%% file: short.tex
\documentclass[twocolumn,english]{revtex4-2}
\usepackage[T1]{fontenc}
\usepackage[latin9]{inputenc}
\usepackage{color}
\usepackage{babel}
\usepackage{array}
\usepackage{multirow}
\usepackage{amsmath}
\usepackage{amssymb}
\usepackage{mathdots}
\usepackage{graphicx}
\usepackage[unicode=true,pdfusetitle,
 bookmarks=true,bookmarksnumbered=false,bookmarksopen=false,
 breaklinks=false,pdfborder={0 0 0},pdfborderstyle={},backref=false,colorlinks=true]
 {hyperref}
\hypersetup{
 linkcolor=blue,citecolor=blue,urlcolor=black}

\makeatletter

\providecommand{\tabularnewline}{\\}

\makeatother

\begin{document}
\title{Exact overlaps for ''all'' integrable matrix product states of rational
spin chains}
\author{Tamas Gombor}
\affiliation{MTA-ELTE \textquotedblleft Momentum\textquotedblright{} Integrable
Quantum Dynamics Research Group, Eötvös Loránd University, Pázmány
Péter sétány 1/A, 1117 Budapest, Hungary}
\affiliation{HUN-REN Wigner Research Centre for Physics, Konkoly-Thege Miklós u. 29-33,
1121 Budapest, Hungary}
\begin{abstract}
The overlaps between integrable matrix product states (MPS) and Bethe
states are important in both the non-equilibrium statistical physics
and the AdS/CFT duality. We present the general MPS overlap formula.
The result is a product of a ratio of Gaudin determinants and a prefactor.
The Gaudin determinants depend on the spin chain but not on the MPS.
The MPS dependent prefactor is given for all integrable MPS of the
$\mathfrak{gl}_{N}$, $\mathfrak{o}_{N}$ and $\mathfrak{sp}_{N}$
symmetric spin chains with arbitrary representations.
\end{abstract}
\maketitle

\section{Introduction}

Over the past ten years, intensive research has been carried out on
the topic of integrable boundary states. The overlaps between boundary
and Bethe states have important applications in two areas of theoretical
physics: non-equilibrium dynamics in statistical physics and the AdS/CFT
duality.

A major breakthrough in the last decade has been the laboratory simulation
of integrable statistical physics models \citep{Hackerm_ller_2010,Schneider2012FermionicTA,PhysRevLett.122.090601}.
A common feature is the experimental study of so-called quantum quenches,
where certain initial states (ground state of the pre-quenched system)
are prepared and the resulting finite-time dynamics are monitored.
This launched an intensive theoretical investigation of the non-equilibrium
dynamics of integrable models \citep{Calabrese_2016,Bastianello_2022}.
These studies are based on the so-called Quench Action method, which
allows the calculation of the long-term equilibrium state and the
possibility to study the time evolution of specific initial states
\citep{Wouters_2014,Caux:2013ra,Essler:2016ufo}. The Quench Action
method is based on explicit knowledge of \emph{the overlaps between
the initial state and the Bethe eigenstates}. This method has been
successfully applied to a number of different models \citep{PhysRevLett.116.070408,Piroli:2018ksf,Rylands:2022naf,Rylands:2022gev},
but the investigated initial states were the so-called two-site product
states which have essentially minimal entanglement.

In parallel, the investigation of the AdS/CFT duality with defects
by integrability methods was started. First, it was found that the
one-point functions can be traced back to \emph{overlaps of certain
matrix product states (MPS) and Bethe states} for the so-called D3-D5
type domain wall defect of the $\mathcal{N}=4$ super Yang-Mills theory
($\mathcal{N}=4$ SYM) \citep{deLeeuw:2015hxa,Buhl-Mortensen:2015gfd}.
These overlap formulas, which give the one-point functions at weak
coupling, have been conjectured and verified numerically \citep{DeLeeuw:2018cal}.
Later, boundary states at finite coupling were defined \citep{Gombor:2020kgu,Komatsu:2020sup},
and their corresponding overlap functions give the finite coupling
one-point functions in the asymptotic limit \citep{Gombor:2020auk,Buhl-Mortensen:2017ind}.
This domain wall setup was also generalized for the ABJM theory \citep{Kristjansen:2021abc,Gombor:2022aqj}.
Similar boundary states have been found for other defects. In the
presence of giant graviton some correlation functions are also given
by boundary state overlaps in the $\mathcal{N}=4$ SYM \citep{Jiang:2019zig}
and in the ABJM \citep{Yang:2021hrl}. Boundary state overlaps also
appear for the one-point functions in the presence of Wilson- and
't Hooft lines \citep{Kristjansen:2023ysz,Jiang:2023cdm}. Most recently,
the integrable boundary states and their overlaps emerged in the investigation
of the Coulomb branch of $\mathcal{N}=4$ SYM \citep{Ivanovskiy:2024vel}.
The integrable conformal defects in $\mathcal{N}=4$ SYM was also
classified \citep{deLeeuw:2024qki}. The physical significance of
these defects is that they violate conformal \citep{Ivanovskiy:2024vel}
or super-symmetry \citep{deLeeuw:2016ofj} in certain configurations,
which can lead to more realistic theories. In these models, the role
of the boundary state overlaps is important.

Due to the above statistical physical and AdS/CFT duality applications,
the research on the overlaps of integrable boundary and Bethe states
has been initiated. The definition of the integrable boundary states
appeared first in the context of 2d quantum field theories \citep{Ghoshal:1993tm}
and later was generalized to spin chains \citep{Piroli:2017sei,Pozsgay:2018dzs}.
The first important observation on the overlaps was that they are
proportional to the ratio of Gaudin determinants \citep{Brockmann_2014}.
The Gaudin determinants are universal, i.e., they do not depend on
the boundary state. The boundary state dependent part of the overlap
function is the prefactor before the ratio. Later this observation
was used to conjecture overlap formulas \citep{deLeeuw:2016umh,DeLeeuw:2018cal,Pozsgay_Overlap}.
There have been only a few proofs \citep{Brockmann_2014,Foda:2015nfk,Jiang:2020sdw},
mainly concentrated on spin 1/2 XXX or XXZ spin chains. It was not
clear how these methods could be generalized to general spins or higher
rank symmetries. Later, a method based on the algebraic Bethe Ansatz
was developed to determine and prove the overlap formulas of $\mathfrak{gl}_{N}$
symmetric spin chains for any representations and any two-site product
states \citep{Gombor:2021uxz,Gombor:2021hmj,Gombor:2023bez}. Although
these results are extremely general, they do not say anything about
MPS overlaps. In terms of the applications, the MPS overlaps are particularly
important. For quantum quenches, the ground state of the pre-quench
system can be approximated better with an MPS than with a two-site
product state. Even in the AdS/CFT duality, the most interesting defects
correspond to MPS boundary states.

In this letter we present the general MPS overlap formula (\ref{eq:overlap}).
Every integrable MPS corresponds to a $K$-matrix which is a representation
of a reflection algebra. One can define a series of commuting matrices
($F$-operators) from the components of the $K$-matrix. The spectrum
of the $F$-operators completely defines the MPS overlap formula.
The definition of the $F$-operators depends only on the reflection
algebra. In the Supplemental Materials we give the definitions of
the $F$-operators for every reflection algebras of the $\mathfrak{gl}_{N}$,
$\mathfrak{o}_{N}$ or $\mathfrak{sp}_{N}$ symmetric spin chains.
These results give the overlaps for all integrable MPS of the $\mathfrak{gl}_{N}$,
$\mathfrak{o}_{N}$ or $\mathfrak{sp}_{N}$ symmetric spin chains
with arbitrary representations.

\section{Integrable spin chains and matrix product states}

\subsection{Definitions for integrable spin chains}

In this letter we do not focus on a particular model but we work with
a wide class of integrable spin chains. A spin chain is given by a
Hamiltonian $H$. The $H$ acts on a Hilbert space $\mathcal{H}$
with a tensor product form $\mathcal{H}=\left(\mathbb{C}^{d}\right)^{J}$
where $d$ is the dimension of the one-site Hilbert space and $J$
is the length of the spin chain. The model is integrable if there
exists a generating function $\mathcal{T}(u)$ of the commuting charges
which commute with the Hamiltonian
\begin{equation}
\left[H,\mathcal{T}(u)\right]=0.\label{eq:HT}
\end{equation}
The generating function $\mathcal{T}(u)$ is the transfer matrix.
Let us introduce an auxiliary space $\mathbb{C}^{N}$. For integrable
spin chains with periodic boundary condition the transfer matrix is
\begin{equation}
\mathcal{T}(u)=\sum_{j=1}^{N}T_{j,j}(u),
\end{equation}
where $T_{i,j}(u)$ are the entries of the monodromy matrix for $i,j=1,\dots,N$.
We also introduce the matrix notation in the auxiliary space $T(u)=\sum_{i,j}e_{i,j}\otimes T_{i,j}(u)\in\mathrm{End}(\mathbb{C}^{N}\otimes\mathcal{H})$,
where $e_{i,j}$ are the unit matrices of $\mathbb{C}^{N}$ which
have the components $\left(e_{i,j}\right)_{a,b}=\delta_{i,a}\delta_{j,b}$.
The monodromy matrix satisfies the famous $RTT$-relation
\begin{equation}
R_{1,2}(u-v)T_{1}(u)T_{2}(v)=T_{2}(v)T_{1}(u)R_{1,2}(u-v),\label{eq:RTT}
\end{equation}
where the $R(u)\in\mathrm{End}(\mathbb{C}^{N}\otimes\mathbb{C}^{N})$
is the R-matrix and the subscripts $1,2$ label two copies of the
auxiliary space. The $RTT$-relation has two consequences: the R-matrix
has to satisfy the famous Yang-Baxter (YB) equation
\begin{equation}
R_{1,2}(u-v)R_{1,3}(u)R_{2,3}(v)=R_{2,3}(v)R_{1,3}(u)R_{1,2}(u-v),
\end{equation}
and the transfer matrix generates commuting operators
\begin{equation}
\left[\mathcal{T}(u),\mathcal{T}(v)\right]=0.
\end{equation}

In this letter we investigate rational spin chains which have Lie-algebra
symmetry $\mathfrak{gl}_{N}$, $\mathfrak{o}_{N}$ or $\mathfrak{sp}_{2n}$.
The $\mathfrak{gl}(N)$ symmetric R-matrix is
\begin{equation}
R(u)=\mathbf{1}+\frac{1}{u}\mathbf{P},\quad\mathbf{P}=\sum_{i,j=1}^{N}e_{i,j}\otimes e_{j,i}.
\end{equation}
The $\mathfrak{o}_{N}$ and $\mathfrak{sp}_{N}$ symmetric R-matrices
can be written as
\begin{equation}
R(u)=\mathbf{1}+\frac{1}{u}\mathbf{P}-\frac{1}{u+\kappa_{N}}\mathbf{Q},\quad\mathbf{Q}=\sum_{i,j=1}^{N}\theta_{i}\theta_{j}e_{i,j}\otimes e_{\bar{i},\bar{j}},
\end{equation}
where $\bar{i}=N+1-i$ and $\kappa_{N}=\frac{N\mp2}{2}$. The minus
or plus signs correspond to the $\mathfrak{o}_{N}$ or $\mathfrak{sp}_{N}$
algebras, respectively. For $\mathfrak{o}_{N}$ $\theta_{i}=+1$ and
for $\mathfrak{sp}_{N}$ 
\begin{equation}
\theta_{i}=\begin{cases}
+1, & i=1,\dots,\frac{N}{2},\\
-1, & i=\frac{N}{2}+1,\dots,N.
\end{cases}
\end{equation}
The monodromy matrices are representations of the Yangian algebras
$Y(\mathfrak{g})$ \citep{Molev:1994rs}, where $\mathfrak{g}=\mathfrak{gl}_{N},\mathfrak{o}_{N}$
or $\mathfrak{sp}_{2n}$. 

\begin{figure}
\begin{centering}
\includegraphics[width=0.9\columnwidth]{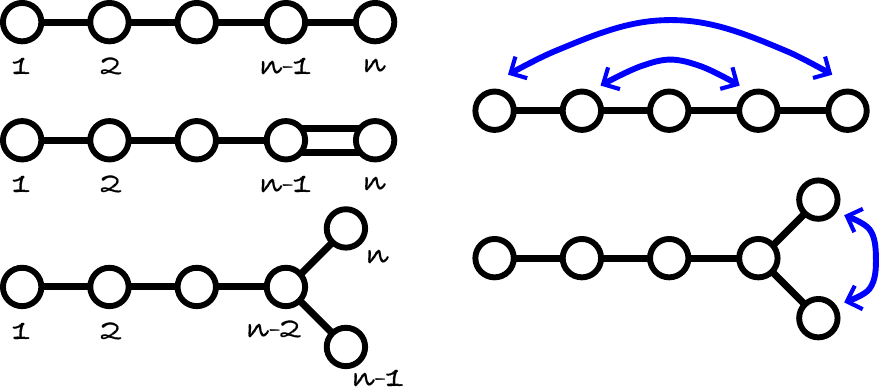}
\par\end{centering}
\caption{Dynkin diagrams (first row $\mathfrak{gl}_{n+1}$, second row $\mathfrak{o}_{2n+1}$
or $\mathfrak{sp}_{2n}$, third row $\mathfrak{o}_{2n}$) and the
non-trivial involutions which correspond to achiral pair structures.}

\label{fig:Dynkin}
\end{figure}

Let us introduce the eigenvalues and eigenvectors of the transfer
matrix
\begin{equation}
\mathcal{T}(u)|\bar{u}\rangle=\tau(u|\bar{u})|\bar{u}\rangle,
\end{equation}
where $|\bar{u}\rangle$ is the Bethe vector and $\bar{u}\equiv\left\{ \bar{u}^{j}\right\} _{j=1}^{n}$
denotes the set of Bethe roots $\bar{u}^{j}=\left\{ u_{k}^{j}\right\} _{k=1}^{r_{j}}$.
The $n$ is the rank of the symmetry algebra ($n$ for $\mathfrak{gl}_{n+1}$,
$\mathfrak{o}_{2n+1}$, $\mathfrak{sp}_{2n}$ and $\mathfrak{o}_{2n}$).
The Bethe roots should satisfy the Bethe Ansatz equations \citep{OGIEVETSKY1986360}

\begin{equation}
\alpha_{a}(u_{k}^{a})\prod_{b=1}^{n}\prod_{l=1}^{r_{b}}\frac{u_{k}^{a}-u_{l}^{b}-\frac{i}{2}C_{a,b}}{u_{k}^{a}-u_{l}^{b}+\frac{i}{2}C_{a,b}}=-1,
\end{equation}
for $a=1,\dots,n$ and $k=1,\dots,r_{a}$. The $C_{a,b}$ are the
inner products of the simple roots of the corresponding algebra \footnote{We give the explicit form in the Supplemental Materials.},
the functions $\alpha_{a}(u)$ depends on the representations of the
Hilbert space and the inhomogeneities. A set of the Bethe roots $\bar{u}^{a}$
corresponds to a node $a$ of the Dynkin diagram for $a=1,\dots,n$,
see figure \ref{fig:Dynkin}. 

\subsection{Integrable matrix product states}

For even $J$ let us define the following matrix product state (MPS)
\begin{multline}
\langle\mathrm{MPS}|=\\
\sum_{i_{1},\dots,i_{J}}\mathrm{Tr}\left[\psi_{i_{J-1},i_{J}}^{(J/2)}\dots\psi_{i_{1},i_{2}}^{(1)}\right]\langle i_{1},i_{2},\dots,i_{J-1},i_{J}|,\label{eq:2site}
\end{multline}
where $\psi_{i,j}^{(k)}\in\mathrm{End}(\mathbb{C}^{d_{B}})$ for $i,j=1,\dots,d$,
$k=1,\dots,J/2$ and $d_{B}$ is the boundary (or bond) dimension.
The $\psi_{i,j}^{(k)}$ matrices can be different (i.e. $\psi_{i,j}^{(k)}\neq\psi_{i,j}^{(l)}$)
for inhomogeneous chains but they are given by the same functions
as $\psi_{i,j}^{(k)}\equiv\psi_{i,j}(\theta_{k})$, where $\theta_{k}$-s
are the inhomogeneities. For homogenous chains it can happen that
$\psi_{i,j}^{(k)}=\omega_{j}\omega_{i}$ for which the MPS can be
also defined for even or odd length:
\begin{equation}
\langle\mathrm{MPS}|=\sum_{i_{1},\dots,i_{J}}\mathrm{Tr}\left[\omega_{i_{J}}\dots\omega_{i_{1}}\right]\langle i_{1},i_{2},\dots,i_{J-1},i_{J}|.\label{eq:1site}
\end{equation}
We also define states by ''cutting the trace''
\begin{multline}
\langle\Psi_{\alpha,\beta}|=\\
\sum_{i_{1},\dots,i_{J}}\left(\psi_{i_{J-1},i_{J}}^{(J/2)}\dots\psi_{i_{1},i_{2}}^{(1)}\right)_{\alpha,\beta}\langle i_{1},i_{2},\dots,i_{J-1},i_{J}|,
\end{multline}
where $\alpha,\beta=1,\dots,d_{B}$ are indices of the boundary space.
The boundary state is integrable if it satisfies the $KT$-relation
\begin{equation}
\sum_{k,\gamma}K_{i,k}^{\alpha,\gamma}(u)\langle\Psi_{\gamma,\beta}|T_{k,j}(u)=\sum_{k,\gamma}\langle\Psi_{\alpha,\gamma}|\widehat{T}_{i,k}(-u)K_{k,j}^{\gamma,\beta}(u),\label{eq:KT}
\end{equation}
where the coefficients $K_{i,j}^{\alpha,\beta}(u)\in\mathbb{C}$ define
the K-matrix $\mathbf{K}_{i,j}(u)=\sum_{\alpha,\beta}K_{i,j}^{\alpha,\beta}(u)e_{\alpha,\beta}^{B}\in\mathrm{End}(\mathbb{C}^{d_{B}})$,
$\mathbf{K}(u)=\sum_{i,j}e_{i,j}\otimes\mathbf{K}_{i,j}(u)\in\mathrm{End}(\mathbb{C}^{N}\otimes\mathbb{C}^{d_{B}})$.
We used the elementary matrices $(e_{\alpha,\beta}^{B})_{a,b}=\delta_{\alpha,a}\delta_{\beta,b}$
of the boundary space $\mathbb{C}^{d_{B}}$. We also define the notation
$\langle\Psi|\in\mathcal{H}^{*}\otimes\mathrm{End}(\mathbb{C}^{d_{B}})$
which is a covector in the Hilbert space and a matrix in the boundary
space: $\langle\Psi|=\sum_{\alpha,\beta}\langle\Psi_{\alpha,\beta}|\otimes e_{\alpha,\beta}^{B}$.
Using these matrix notations in the boundary space the $KT$-relation
has the following compact form
\begin{equation}
\mathbf{K}(u)\langle\Psi|T(u)=\langle\Psi|\widehat{T}(-u)\mathbf{K}(u).\label{eq:KT-1}
\end{equation}
Here the tensors $\mathbf{K}(u)$, $\langle\Psi|$ or $T(u)$ act
trivially on the Hilbert-, the auxiliary- or the boundary spaces,
respectively. 

There are two types of $KT$-relation: the uncrossed $KT$-relation
where $\widehat{T}=T$ and the crossed $KT$-relation where 
\begin{equation}
\sum_{k=1}^{N}\widehat{T}_{k,i}(u)T_{k,j}(u)=\lambda_{1}(u)\lambda_{1}(-u)\delta_{i,j}\mathbf{1}.\label{eq:Tch}
\end{equation}
Here we used the pseudo-vacuum eigenvalue $T_{1,1}(u)|0\rangle=\lambda_{1}(u)|0\rangle$
\footnote{The pseudo-vacuum is the Bethe vector without Bethe roots, i.e., $r_{1}=r_{2}=\dots=r_{n}=0$.}.
Assuming that the $K$-marix is invertible, the condition
\begin{equation}
\langle\mathrm{MPS}|\mathcal{T}(u)=\langle\mathrm{MPS}|\widehat{\mathcal{T}}(-u),\label{eq:intcond}
\end{equation}
is following from the $KT$-relation.

\begin{table}
\begin{centering}
\begin{tabular}{|c|c|c|c|c|}
\hline 
$\mathfrak{g}$ & $\mathfrak{h}$ & KT-relation & P.s. & $n_{+}$\tabularnewline
\hline 
\hline 
\multirow{3}{*}{$\mathfrak{gl}_{N}$} & $\mathfrak{o}_{N}$ & crossed & \textcolor{red}{chiral} & $N-1$\tabularnewline
\cline{2-5} \cline{3-5} \cline{4-5} \cline{5-5} 
 & $\mathfrak{sp}_{N}$ & crossed & \textcolor{red}{chiral} & $N-1$\tabularnewline
\cline{2-5} \cline{3-5} \cline{4-5} \cline{5-5} 
 & $\mathfrak{gl}_{M}\oplus\mathfrak{gl}_{N-M}$ & uncrossed & \textcolor{blue}{achiral} & $\left\lfloor \frac{N}{2}\right\rfloor $\tabularnewline
\hline 
\hline 
$\mathfrak{o}_{2n+1}$ & $\mathfrak{o}_{M}\oplus\mathfrak{o}_{2n+1-M}$ & uncrossed & \textcolor{red}{chiral} & $n$\tabularnewline
\hline 
\hline 
\multirow{2}{*}{$\mathfrak{sp}_{2n}$} & $\mathfrak{sp}_{2m}\oplus\mathfrak{sp}_{2n-2m}$ & uncrossed & \textcolor{red}{chiral} & $n$\tabularnewline
\cline{2-5} \cline{3-5} \cline{4-5} \cline{5-5} 
 & $\mathfrak{gl}_{n}$ & uncrossed & \textcolor{red}{chiral} & $n$\tabularnewline
\hline 
\hline 
\multirow{3}{*}{$\mathfrak{o}_{2n=4k}$} & $\mathfrak{o}_{2m}\oplus\mathfrak{o}_{2n-2m}$ & uncrossed & \textcolor{red}{chiral} & $n$\tabularnewline
\cline{2-5} \cline{3-5} \cline{4-5} \cline{5-5} 
 & $\mathfrak{o}_{2m+1}\oplus\mathfrak{o}_{2n-2m-1}$ & uncrossed & \textcolor{blue}{achiral} & $n-1$\tabularnewline
\cline{2-5} \cline{3-5} \cline{4-5} \cline{5-5} 
 & $\mathfrak{gl}_{n}$ & uncrossed & \textcolor{red}{chiral} & $n$\tabularnewline
\hline 
\hline 
\multirow{3}{*}{$\mathfrak{o}_{2n=4k+2}$} & $\mathfrak{o}_{2m}\oplus\mathfrak{o}_{2n-2m}$ & uncrossed & \textcolor{blue}{achiral} & $n-1$\tabularnewline
\cline{2-5} \cline{3-5} \cline{4-5} \cline{5-5} 
 & $\mathfrak{o}_{2m+1}\oplus\mathfrak{o}_{2n-2m-1}$ & uncrossed & \textcolor{red}{chiral} & $n$\tabularnewline
\cline{2-5} \cline{3-5} \cline{4-5} \cline{5-5} 
 & $\mathfrak{gl}_{n}$ & uncrossed & \textcolor{blue}{achiral} & $n-1$\tabularnewline
\hline 
\end{tabular}
\par\end{centering}
\caption{The reflections algebras $Y(\mathfrak{g},\mathfrak{h})$ and the corresponding
pair structures.}

\label{tab:twY}
\end{table}

Another consequence of the $KT$-relation is that the $K$-matrix
satisfies the reflection equation
\begin{multline}
R_{1,2}(u-v)\mathbf{K}_{1}(-u)\bar{R}_{1,2}(u+v)\mathbf{K}_{2}(-v)=\\
\mathbf{K}_{2}(-v)\bar{R}_{1,2}(u+v)\mathbf{K}_{1}(-u)R_{1,2}(u-v),
\end{multline}
where
\begin{equation}
\bar{R}_{1,2}(u)=\begin{cases}
R_{1,2}(u), & \text{for the uncrossed }KT,\\
R_{1,2}^{t_{2}}(-u), & \text{for the crossed }KT.
\end{cases}
\end{equation}
The reflection equation defines reflection algebra with the generators
$\mathbf{K}(u)$. Let us take the series expansion
\begin{equation}
\mathbf{K}(u)=\mathcal{U}\otimes\mathbf{1}+u^{-1}\sum_{i,j=1}^{N}e_{i,j}\otimes\mathbf{k}_{i,j}^{(1)}+\mathcal{O}(u^{-2}),\label{eq:Femb}
\end{equation}
where $\mathcal{U}\in\mathrm{Aut}(\mathbb{C}^{N})$ and $\mathbf{k}_{i,j}^{(1)}\in\mathrm{End}(\mathbb{C}^{d_{B}})$.
The matrices $\mathbf{k}_{i,j}^{(1)}$ are generates Lie-subalgebra
$\mathfrak{h}$ \citep{Gombor:2019bun}, and the $K$-matrix is a
representation of the twisted Yangian $Y(\mathfrak{g},\mathfrak{h})$
\citep{Molev:1997wp,MOLEV:2002,Regelskis:2017,Regelskis:2019}. The
classification of the twisted Yangians is shown in table \ref{tab:twY}.

\section{Exact overlaps of integrable MPS}

\subsection{Pair structures}

Now we continue with the overlaps between the integrable MPS-s and
the on-shell Bethe states. The integrability condition (\ref{eq:intcond})
is equivalent to non-trivial selection rules for the on-shell overlaps,
namely, the on-shell overlap is non-vanishing only when the Bethe
roots have pair structures, i.e. $\bar{u}^{j}=-\bar{u}^{\rho(j)}$
for $j=1,\dots,n$. The $\rho$ is an involution of the Dynkin-diagram.
If $\rho$ is trivial, i.e., $\rho(j)=j$ for every $j$ then we have
\emph{chiral} pair structure. The pair structure is \emph{achiral}
for non-trivial $\rho$. We have two possibilities for non-trivial
$\rho$: for $\mathfrak{gl}_{N}$ $\rho(j)=N-j$; and for $\mathfrak{o}_{2n}$
$\rho(n-1)=n$, $\rho(n)=n-1$ and $\rho(j)=j$ for $j<n-1$. For
$\mathfrak{o}_{2n+1}$ and $\mathfrak{so}_{2n}$ there is no non-trivial
involution therefore only the chiral pair structure is possible. Every
twisted Yangian has a specific pair structure which is summarized
in table \ref{tab:twY}. Integrable MPS with a given chirality can
only exist for spin chains where the following conditions are satisfied
$\alpha_{j}(u)\alpha_{\rho(j)}(-u)=1$ for $j=1,\dots,n$.

Also, a set of Bethe roots $\bar{u}^{j}$ is chiral if the condition
$\rho(j)=j$ holds. For the sake of simplicity, we focus on the states
where every chiral set has an even number of elements. We cover the
odd cases separately in the Supplemental Materials. The pair structures
halve the degrees of freedom, i.e. half of the Bethe roots can be
expressed from the other half. More precisely, the full set has a
decomposition $\bar{u}=\bar{u}^{+}\cup\bar{u}^{-}$ for which $\bar{u}^{+},\bar{u}^{-}$
are disjoint subsets and $\bar{u}^{-}=-\bar{u}^{+}$ in the pair structure
limit. Let $n_{+}$ be the number of sets in $\bar{u}^{+}$. The explicit
values of $n_{+}$ are listed in table \ref{tab:twY}.

\subsection{Overlap formula }

The on-shell overlaps are
\begin{equation}
\frac{\langle\mathrm{MPS}|\bar{u}\rangle}{\sqrt{\langle\bar{u}|\bar{u}\rangle}}={\color{blue}\left[\sum_{k=1}^{d_{B}}\beta_{k}\prod_{j=1}^{n_{+}}\tilde{\mathcal{F}}_{k}^{(j)}(\bar{u}^{+,j})\right]}{\color{red}\sqrt{\frac{\det G^{+}}{\det G^{-}}}}.\label{eq:overlap}
\end{equation}
The overlap formula has two parts. The ratio of Gaudin determinants
(red) does not depend on MPS. They depend only on the Bethe ansatz
equations and the pair structure. The Gaudin determinants are defined
in the Supplementary material but they are also found in several previous
papers \citep{Brockmann_2014,Gombor:2021hmj,Gombor:2023bez}. 

The prefactor (blue) is the MPS dependent part. We introduced shorthand
notation $\tilde{\mathcal{F}}_{k}^{(j)}(\bar{v})=\prod_{v_{l}\in\bar{v}}\tilde{\mathcal{F}}_{k}^{(j)}(v_{l})$
where $\tilde{\mathcal{F}}_{k}^{(j)}(u)$-s are single-variable functions
for $j=1,\dots,n_{+}$ and $k=1,\dots,d_{B}$. These functions $\tilde{\mathcal{F}}_{k}^{(j)}(u)$
depend only on the $K$-matrix, but not on the concrete representation
of the Hilbert space. For every twisted Yangian $Y(\mathfrak{g},\mathfrak{h})$
there is a commuting subalgebra which are generated by $\mathbf{F}^{(j)}(u)$
for $j=1,\dots,n_{+}$. The $F$-operators $\mathbf{F}^{(j)}(u)$
can be expressed by the components of the $K$-matrix. Since these
are commuting operators they can be diagonalized simultaneously and
the eigenvalues give $\tilde{\mathcal{F}}_{k}^{(j)}(u)$. 

The prefactor also contains the coefficients $\beta_{k}\in\mathbb{C}$
for $k=1,\dots,d_{B}$. They are the eigenvalues of the pseudo-vacuum
overlap operator $\mathbf{B}=\langle\Psi|0\rangle$ which commutes
with the $F$-operators. 

\subsection{Example $Y(\mathfrak{gl}_{N},\mathfrak{o}_{N})$}

Here we give the concrete functions $\tilde{\mathcal{F}}_{k}^{(j)}(u)$
for the reflection algebra $Y(\mathfrak{gl}_{N},\mathfrak{o}_{N})$.
The non-vanishing overlaps require chiral pair structure, i.e., there
exist decompositions $\bar{u}^{s}=\bar{u}^{+,s}\cup\bar{u}^{-,s}$
for $s=1,\dots,N-1$. In the pair structure limit $\bar{u}^{-,s}=-\bar{u}^{+,s}$.
For chiral pair structure $n_{+}=N-1$ therefore we have $N-1$ $F$-operators. 

First we define a series of nested $K$-matrices 
\begin{equation}
\mathbf{K}_{a,b}^{(k+1)}(u)=\mathbf{K}_{a,b}^{(k)}(u)-\mathbf{K}_{a,k}^{(k)}(u)\left[\mathbf{K}_{k,k}^{(k)}(u)\right]^{-1}\mathbf{K}_{k,b}^{(k)}(u),\label{eq:kdef}
\end{equation}
with the initial condition $\mathbf{K}^{(1)}(u)\equiv\mathbf{K}(u)$.
The $F$-operators are given by
\begin{equation}
\mathbf{F}^{(k)}(u)=\left[\mathbf{K}_{k,k}^{(k)}(u)\right]^{-1}\mathbf{K}_{k+1,k+1}^{(k+1)}(u),\label{eq:fdef}
\end{equation}
for $k=1,\dots,N-1$. They generate a commuting subalgebra and they
also commute with the operator $\mathbf{B}=\langle\Psi|0\rangle$,
i.e.
\begin{equation}
\left[\mathbf{F}^{(k)}(u),\mathbf{F}^{(l)}(v)\right]=\left[\mathbf{F}^{(k)}(u),\mathbf{B}\right]=0,
\end{equation}
for $k,l=1,\dots,N-1$. They can be diagonalized simultaneously
\begin{equation}
\begin{split}\mathbf{F}^{(k)}(u) & =\mathbf{A}\mathrm{diag}(\mathcal{F}_{1}^{(k)}(u),\dots,\mathcal{F}_{d_{B}}^{(k)}(u))\mathbf{A}^{-1},\\
\mathbf{B} & =\mathbf{A}\mathrm{diag}(\beta_{1},\dots,\beta_{d_{B}})\mathbf{A}^{-1},
\end{split}
\end{equation}
where $\mathbf{A}$ is an invertible matrix in the boundary space.
Finally, the $\tilde{\mathcal{F}}_{k}^{(j)}(u)$ functions of the
overlap formula (\ref{eq:overlap}) are
\begin{equation}
\tilde{\mathcal{F}}_{j}^{(k)}(u)=\mathcal{F}_{j}^{(k)}(iu-k/2)\sqrt{\frac{u^{2}}{u^{2}+1/4}}.
\end{equation}

This result gives the overlaps for any representation of the reflection
algebra $Y(\mathfrak{gl}_{N},\mathfrak{o}_{N})$. The definitions
of the $F$-operators depend only on the reflection algebra. In the
Supplementary Material we give the concrete definitions of $\tilde{\mathcal{F}}_{k}^{(j)}(u)$
for all twisted Yangians $Y(\mathfrak{g},\mathfrak{h})$ of table
\ref{tab:twY} case by case.

\section{How to apply the formula}

Although the end result is very general, its application is not a
trivial task. In this section, we describe the steps for applying
the formula (\ref{eq:overlap}). A typical case might be that we have
an integrable spin chain with an MPS, and we want to decide whether
the MPS is integrable, and if so, we want to write up the overlap
formula. In this case we need to follow the steps below. We assume
that we know the monodromy matrix which satisfies the $RTT$-relation
(\ref{eq:RTT}) and commutes with the Hamiltonian of our model (\ref{eq:HT}).
For the $\mathfrak{gl}_{N}$ symmetric model we may also need the
inverse monodromy matrix (\ref{eq:Tch}). 

\paragraph{1. Deciding on integrability. }

We can check the integrability of the MPS with the equation (\ref{eq:intcond}).
At this step, note that the spectral parameter of the monodromy matrix
can be shifted ($T(u-a)$ also satisfies the defining equations (\ref{eq:RTT}),
(\ref{eq:HT})) , therefore it can happen that only the shifted equation
\begin{equation}
\langle\mathrm{MPS}|\mathcal{T}(u)=\langle\mathrm{MPS}|\widehat{\mathcal{T}}(-u+b),\label{eq:intcond-1}
\end{equation}
 is satisfied in our initial convention. In this case we only need
to change our convention as $T(u)\to T(u+b/2)$ to obtain the integrability
condition (\ref{eq:intcond}). In the $\mathfrak{gl}_{N}$ case, it
must also be decided whether the uncrossed or the crossed equation
is satisfied (or neither).

\paragraph{2. Finding the K-matrix. }

After we verified the integrability of the MPS (and fixed the convention
for the spectral parameter), we have to find the $K$-matrix $\mathbf{K}(u)$
by solving the $KT$-relation (\ref{eq:intcond}). We can choose as
short spin chain as possible. For the MPS-s which are built from two-site
operators (\ref{eq:2site}) we can use $J=2$ and for the one-site
operator MPS (\ref{eq:1site}) we can use $J=1$. Solving the KT-relation
is a simple task since this equation is linear in the components of
$\mathbf{K}(u)$.

\paragraph{3. Identify the reflection algebra}

Now we have to find the twisted Yangian $Y(\mathfrak{g},\mathfrak{h})$
algebra corresponding to our K-matrix. For this we have to identify
the Lie-subalgebra $\mathfrak{h}$ from the series expansion of the
$K$-matrix (\ref{eq:Femb}). The knowledge of twisted Yangian $Y(\mathfrak{g},\mathfrak{h})$
already selects the pair structure which defines the Gaudin determinants
(see Supplemental Materials).

\paragraph{4. ''Rotations'' of the K-matrix. (if necessary)}

It is possible that the MPS is in the ''wrong direction'' for the
overlap formula. It means the following. The formulas of the $F$-operators
(\ref{eq:kdef}),(\ref{eq:fdef}) assumes that some components of
the $K$-matrix are invertible. But this may not be true for our particular
MPS. However, in all cases there is a $G=\exp(\mathfrak{g})$ transformation
that makes the ''desired'' elements invertible. The group element
$g\in G$ acts on the MPS and the K-matrix in the following way
\begin{equation}
\begin{split}\langle\Psi_{\alpha,\beta}^{g}| & =\langle\Psi_{\alpha,\beta}|\Delta(g),\\
\mathbf{K}^{g}(u) & =\left(\hat{g}^{-1}\otimes\mathbf{1}\right)\mathbf{K}(u)\left(g\otimes\mathbf{1}\right),
\end{split}
\end{equation}
where $\hat{g}=g$ except for the crossed $\mathfrak{gl}_{N}$ when
$\hat{g}=\left(g^{t}\right)^{-1}$. The $\Delta(g)$ is the usual
co-product corresponding to the Hilbert space. The $G$-action preserves
the $KT$-relation. Our task is to find a group element $g(\varphi)=\exp(\varphi x)$
($x\in\mathfrak{g}$,$\varphi\in\mathbb{C}$) for which the definition
of the $F$-operators are well defined. The further steps are performed
on the ''rotated'' K-matrix $\mathbf{K}^{g(\varphi)}$ (if the rotation
is necessary), and we can take the $\varphi\to0$ limit in the final
formula.

\paragraph{5. Calculation of the F-operators}

Since we know the $K$-matrix and the associated reflection algebra
$Y(\mathfrak{g},\mathfrak{h})$, all we have to do is to calculate
the $F$-operators based on the section of $Y(\mathfrak{g},\mathfrak{h})$
in the Supplemental Materials. Their eigenvalues define the functions
$\tilde{\mathcal{F}}_{k}^{(j)}(u)$. 

\section{Conclusions and outlook}

In this letter we gave the exact overlap functions (\ref{eq:overlap})
for every integrable MPS of the $\mathfrak{gl}_{N}$, $\mathfrak{o}_{N}$
and $\mathfrak{sp}_{N}$ symmetric spin chains with arbitrary representations.
For every MPS, there is a corresponding $K$-matrix which is a representation
of a reflection algebra. The $K$-matrix defines a series of commuting
matrices: the $F$-operators. The spectrum of the $F$-operators defines
the MPS overlap formula. In the Supplementary Materials we introduced
the $F$-operators for the reflections algebras of the $\mathfrak{gl}_{N}$,
$\mathfrak{o}_{N}$ and $\mathfrak{sp}_{N}$ symmetric spin chains.

The formulas of this letter have been verified by extensive numerical
tests. The development of precise proofs has been postponed to later
publications.

An interesting future direction could be the extension of the results
to graded spin chains, such as $\mathfrak{gl}(M|N)$ and $\mathfrak{osp}(M|2n)$
symmetric spin chains, or the Hubbard model. The formulas are likely
to be generalizable to the trigonometric cases. 

There exists another method to calculate MPS overlaps based on dressing
formulas \citep{Gombor:2024zru}. The advantage of dressing formulas
is that they can be applied under general twists, but the disadvantage
is that they have to be calculated separately for each representation
of the reflection algebra, unlike the formulas in this letter, which
are true for all representations. 

\paragraph*{Acknowledgments}

The research was supported by the NKFIH grant PD142929 and the János
Bolyai Research Scholarship of the Hungarian Academy of Science.

\bibliographystyle{apsrev4-1}
\bibliography{refs}

\pagebreak

\widetext

\newpage
\begin{center}
\textbf{\large Supplemental Materials: Exact overlaps for ''all'' integrable matrix
product states of rational spin chains}
\end{center}
\setcounter{equation}{0}
\setcounter{figure}{0}
\setcounter{table}{0}
\setcounter{page}{1}
\setcounter{section}{0}
\makeatletter
\renewcommand{\theequation}{S-\arabic{equation}}
\renewcommand{\thefigure}{S-\arabic{figure}}
\renewcommand{\thesection}{S-\Roman{section}}

\input{supp.tex}

\end{document}

%% file: supp.tex
%

\section{Gaudin matrices}

In this section we define the Gaudin determinants $\det G^{\pm}$
of the MPS overlap formula
\begin{equation}
\frac{\langle\mathrm{MPS}|\bar{u}\rangle}{\sqrt{\langle\bar{u}|\bar{u}\rangle}}=\left[\sum_{k=1}^{d_{B}}\beta_{k}\prod_{j=1}^{n_{+}}\tilde{\mathcal{F}}_{k}^{(j)}(\bar{u}^{+,j})\right]\sqrt{\frac{\det G^{+}}{\det G^{-}}}.\label{eq:ov}
\end{equation}
They depend on the Bethe Ansatz equation and the pair structure. In
the following we define them case by case.

The Bethe Ansatz equation are the following.
\begin{itemize}
\item For the $\mathfrak{gl}_{n+1}$ symmetric spin chains the Bethe equations
are
\begin{equation}
e^{i\phi_{k}^{a}}:=\alpha_{a}(u_{k}^{a})\prod_{l=1}^{r_{a}}\frac{u_{k}^{a}-u_{l}^{a}-i}{u_{k}^{a}-u_{l}^{a}+i}\prod_{l=1}^{r_{a-1}}\frac{u_{k}^{a}-u_{l}^{a-1}+\frac{i}{2}}{u_{k}^{a}-u_{l}^{a-1}-\frac{i}{2}}\prod_{l=1}^{r_{a+1}}\frac{u_{k}^{a}-u_{l}^{a+1}+\frac{i}{2}}{u_{k}^{a}-u_{l}^{a+1}-\frac{i}{2}}=-1,\quad a=1,\dots,n.
\end{equation}
\item For the $\mathfrak{o}_{2n+1}$ symmetric spin chains the Bethe equations
are
\begin{equation}
\begin{split}e^{i\phi_{k}^{a}} & :=\alpha_{a}(u_{k}^{a})\prod_{l=1}^{r_{a}}\frac{u_{k}^{a}-u_{l}^{a}-i}{u_{k}^{a}-u_{l}^{a}+i}\prod_{l=1}^{r_{a-1}}\frac{u_{k}^{a}-u_{l}^{a-1}+\frac{i}{2}}{u_{k}^{a}-u_{l}^{a-1}-\frac{i}{2}}\prod_{l=1}^{r_{a+1}}\frac{u_{k}^{a}-u_{l}^{a+1}+\frac{i}{2}}{u_{k}^{a}-u_{l}^{a+1}-\frac{i}{2}}=-1,\quad a=1,\dots,n-1,\\
e^{i\phi_{k}^{n}} & :=\alpha_{n}(u_{k}^{n})\prod_{l=1}^{r_{n}}\frac{u_{k}^{n}-u_{l}^{n}-i/2}{u_{k}^{n}-u_{l}^{n}+i/2}\prod_{l=1}^{r_{n-1}}\frac{u_{k}^{n}-u_{l}^{n-1}+\frac{i}{2}}{u_{k}^{n}-u_{l}^{n-1}-\frac{i}{2}}=-1.
\end{split}
\end{equation}
\item For the $\mathfrak{sp}_{2n}$ symmetric spin chains the Bethe equations
are
\begin{equation}
\begin{split}e^{i\phi_{k}^{a}} & :=\alpha_{a}(u_{k}^{a})\prod_{l=1}^{r_{a}}\frac{u_{k}^{a}-u_{l}^{a}-i}{u_{k}^{a}-u_{l}^{a}+i}\prod_{l=1}^{r_{a-1}}\frac{u_{k}^{a}-u_{l}^{a-1}+\frac{i}{2}}{u_{k}^{a}-u_{l}^{a-1}-\frac{i}{2}}\prod_{l=1}^{r_{a+1}}\frac{u_{k}^{a}-u_{l}^{a+1}+\frac{i}{2}}{u_{k}^{a}-u_{l}^{a+1}-\frac{i}{2}}=-1,\quad a=1,\dots,n-2,\\
e^{i\phi_{k}^{n-1}} & :=\alpha_{n-1}(u_{k}^{n-1})\prod_{l=1}^{r_{n-1}}\frac{u_{k}^{n-1}-u_{l}^{n-1}-i}{u_{k}^{n-1}-u_{l}^{n-1}+i}\prod_{l=1}^{r_{n-2}}\frac{u_{k}^{n-1}-u_{l}^{n-2}+\frac{i}{2}}{u_{k}^{n-1}-u_{l}^{n-2}-\frac{i}{2}}\prod_{l=1}^{r_{n}}\frac{u_{k}^{n-1}-u_{l}^{n}+i}{u_{k}^{n-1}-u_{l}^{n}-i}=-1,\\
e^{i\phi_{k}^{n}} & :=\alpha_{n}(u_{k}^{n})\prod_{l=1}^{r_{n}}\frac{u_{k}^{n}-u_{l}^{n}-2i}{u_{k}^{n}-u_{l}^{n}+2i}\prod_{l=1}^{r_{n-1}}\frac{u_{k}^{n}-u_{l}^{n-1}+i}{u_{k}^{n}-u_{l}^{n-1}-i}=-1.
\end{split}
\end{equation}
\item For the $\mathfrak{o}_{2n}$ symmetric spin chains the Bethe equations
are
\begin{equation}
\begin{split}e^{i\phi_{k}^{a}} & :=\alpha_{a}(u_{k}^{a})\prod_{l=1}^{r_{a}}\frac{u_{k}^{a}-u_{l}^{a}-i}{u_{k}^{a}-u_{l}^{a}+i}\prod_{l=1}^{r_{a-1}}\frac{u_{k}^{a}-u_{l}^{a-1}+\frac{i}{2}}{u_{k}^{a}-u_{l}^{a-1}-\frac{i}{2}}\prod_{l=1}^{r_{a+1}}\frac{u_{k}^{a}-u_{l}^{a+1}+\frac{i}{2}}{u_{k}^{a}-u_{l}^{a+1}-\frac{i}{2}}=-1,\quad a=1,\dots,n-3,\\
e^{i\phi_{k}^{n-2}} & :=\alpha_{n-2}(u_{k}^{n-2})\prod_{l=1}^{r_{n-2}}\frac{u_{k}^{n-2}-u_{l}^{n-2}-i}{u_{k}^{n-2}-u_{l}^{n-2}+i}\prod_{l=1}^{r_{n-3}}\frac{u_{k}^{n-2}-u_{l}^{n-3}+\frac{i}{2}}{u_{k}^{n-2}-u_{l}^{n-3}-\frac{i}{2}}\times\\
 & \times\prod_{l=1}^{r_{n-1}}\frac{u_{k}^{n-2}-u_{l}^{n-1}+\frac{i}{2}}{u_{k}^{n-2}-u_{l}^{n-1}-\frac{i}{2}}\prod_{l=1}^{r_{n}}\frac{u_{k}^{n-2}-u_{l}^{n}+\frac{i}{2}}{u_{k}^{n-2}-u_{l}^{n}-\frac{i}{2}}=-1,\\
e^{i\phi_{k}^{n-1}} & :=\alpha_{n-1}(u_{k}^{n-1})\prod_{l=1}^{r_{n-1}}\frac{u_{k}^{n-1}-u_{l}^{n-1}-i}{u_{k}^{n-1}-u_{l}^{n-1}+i}\prod_{l=1}^{r_{n-2}}\frac{u_{k}^{n-1}-u_{l}^{n-2}+\frac{i}{2}}{u_{k}^{n-1}-u_{l}^{n-2}-\frac{i}{2}}=-1,\\
e^{i\phi_{k}^{n}} & :=\alpha_{n}(u_{k}^{n})\prod_{l=1}^{r_{n}}\frac{u_{k}^{n}-u_{l}^{n}-i}{u_{k}^{n}-u_{l}^{n}+i}\prod_{l=1}^{r_{n-2}}\frac{u_{k}^{n}-u_{l}^{n-2}+\frac{i}{2}}{u_{k}^{n}-u_{l}^{n-2}-\frac{i}{2}}=-1.
\end{split}
\end{equation}
\end{itemize}
The functions $\alpha_{a}(u)$ depends on the representations of the
Hilbert space and the inhomogeneities.

For chiral pair structures $\bar{u}^{s}=-\bar{u}^{s}$ for $s=1,\dots,n$.
We define the quantum numbers $n_{+}=n$ and $r_{s}^{+}=r_{s}/2$
for $s=1,\dots,n$. We introduce new notations for the Bethe roots
$u_{k}^{+,s}\equiv u_{k}^{s}$, $u_{k}^{-,s}\equiv u_{r_{s}+1-k}^{s}$
where $s=1,\dots,n_{+}$ and $k=1,\dots,r_{s}^{+}$. We also define
the following sets $\bar{u}^{\pm,s}=\left\{ u_{k}^{\pm,s}\right\} _{k=1}^{r_{s}^{+}}$
for $s=1,\dots,n_{+}$ and $\bar{u}^{\pm}=\left\{ \bar{u}^{\pm,1},\bar{u}^{\pm,2},\dots,\bar{u}^{\pm,n_{+}}\right\} $.
In the pair structure limit $u_{k}^{-,s}=-u_{k}^{+,s}$ for $s=1,\dots,n_{+}$
and $k=1,\dots,r_{s}^{+}$. Chiral integrable MPS exists only for
the spin chains where the following conditions are satisfied: $\alpha_{s}(u)\alpha_{s}(-u)=1$
for $s=1,\dots,n$. 

The possible achiral pair structures are the following
\begin{itemize}
\item For $\mathfrak{gl}_{2n_{+}+1}$, the achiral pair structure is $\bar{u}^{2n_{+}}=-\bar{u}^{1}$,
$\bar{u}^{2n_{+}-1}=-\bar{u}^{2}$, $\dots$$\bar{u}^{n_{+}+1}=-\bar{u}^{n_{+}}$.
We define the quantum numbers $r_{s}^{+}=r_{s}$ for $s=1,\dots,n_{+}$.
We introduce new notations for the Bethe roots $u_{k}^{+,s}\equiv u_{k}^{s}$
and $u_{k}^{-,s}\equiv u_{k}^{N-s}$ where $s=1,\dots,n_{+}$ and
$k=1,\dots,r_{s}^{+}$. Achiral integrable MPS-s of $\mathfrak{gl}_{2n_{+}+1}$
symmetric spin chains exist only when the following conditions are
satisfied: $\alpha_{s}(u)\alpha_{N-s}(-u)=1$ for $s=1,\dots,n^{+}$. 
\item For $\mathfrak{gl}_{2n_{+}}$, the achiral pair structure is $\bar{u}^{2n_{+}-1}=-\bar{u}^{1}$,
$\bar{u}^{2n_{+}-2}=-\bar{u}^{2}$, $\dots$$\bar{u}^{n_{+}}=-\bar{u}^{n_{+}}$.
We can see that the middle node $\bar{u}^{n_{+}}$ is chiral. We define
the quantum numbers $r_{s}^{+}=r_{s}$ for $s=1,\dots,n_{+}-1$ and
$r_{n_{+}}^{+}=r_{n_{+}}/2$. We introduce new notations for the Bethe
roots $u_{k}^{+,s}\equiv u_{k}^{s}$, $u_{k}^{-,s}\equiv u_{k}^{N-s}$
where $s=1,\dots,n_{+}-1$ and $k=1,\dots,r_{s}^{+}$. For the chiral
node we define $u_{k}^{+,n_{+}}\equiv u_{k}^{n_{+}}$ and $u_{k}^{-,n_{+}}\equiv u_{r_{n_{+}}+1-k}^{n_{+}}$
where $k=1,\dots,r_{n_{+}}^{+}$. Achiral integrable MPS-s of $\mathfrak{gl}_{2n_{+}}$
symmetric spin chains exist only when the following conditions are
satisfied: $\alpha_{s}(u)\alpha_{N-s}(-u)=1$ for $s=1,\dots,n^{+}$. 
\item For $\mathfrak{o}_{2n}$, the achiral pair structure is $\bar{u}^{1}=-\bar{u}^{1}$,
$\bar{u}^{2}=-\bar{u}^{2}$, $\dots$,$\bar{u}^{n-2}=-\bar{u}^{n-2}$
and $\bar{u}^{n}=-\bar{u}^{n-1}$. We can see that the nodes $\bar{u}^{s}$
for $s=1,\dots,n-2$ are chiral. We define the quantum numbers $n_{+}=n-1$,
$r_{s}^{+}=r_{s}/2$ for $s=1,\dots,n-2$ and $r_{n_{+}}^{+}=r_{n-1}=r_{n}$.
We introduce new notations for Bethe roots $u_{k}^{+,s}\equiv u_{k}^{s}$,
$u_{k}^{-,s}\equiv u_{r_{s}+1-k}^{s}$ where $s=1,\dots,n_{+}-1$,
$k=1,\dots,r_{s}^{+}$ and $u_{k}^{+,n_{+}}\equiv u_{k}^{n-1}$, $u_{k}^{-,n_{+}}\equiv u_{k}^{n}$
where $k=1,\dots,r_{n-1}^{+}$. Achiral integrable MPS-s of $\mathfrak{o}_{2n}$
symmetric spin chains exist only when the following conditions are
satisfied: $\alpha_{s}(u)\alpha_{s}(-u)=1$ for $s=1,\dots,n-2$ and
$\alpha_{n-1}(u)\alpha_{n}(-u)=1$. 
\end{itemize}
For all chiral or achiral pair structures we also define the following
sets $\bar{u}^{\pm,s}=\left\{ u_{k}^{\pm,s}\right\} _{k=1}^{r_{s}^{+}}$
where $s=1,\dots,n_{+}$. We also have $\bar{u}^{\pm}=\left\{ \bar{u}^{\pm,1},\bar{u}^{\pm,2},\dots,\bar{u}^{\pm,n_{+}}\right\} $.
In the pair structure limit $u_{k}^{-,s}=-u_{k}^{+,s}$ for $s=1,\dots,n_{+}$
and $k=1,\dots,r_{s}^{+}$.

The expressions $\phi_{k}^{s}$ are depend on the Bethe roots and
they are defined above. These can be grouped according to the pair
structure in the same way as the Bethe roots: $\bar{\phi}^{\pm,s}=\left\{ \phi_{k}^{\pm,s}\right\} _{k=1}^{r_{s}^{+}}$
where $s=1,\dots,n_{+}$ and $\bar{\phi}^{\pm}=\left\{ \bar{\phi}^{\pm,1},\bar{\phi}^{\pm,2},\dots,\bar{\phi}^{\pm,n_{+}}\right\} $.
We can define the components of the Gaudin matrix
\begin{equation}
G_{k,l}^{\mu,\nu}=\frac{\partial\phi_{k}^{\mu}}{\partial u_{l}^{\nu}}.
\end{equation}
The Gaudin matrices have $\sum_{s=1}^{n}r_{s}$ rows and columns and
the rows are labeled by $\nu,l$ and the columns are labeled by $\mu,k$.
According to the decompositions $\bar{u}=\bar{u}^{+}\cup\bar{u}^{-}$
and $\bar{\phi}=\bar{\phi}^{+}\cup\bar{\phi}^{-}$ the Gaudin matrix
has the following block form
\begin{equation}
G=\left(\begin{array}{cc}
A^{++} & A^{+-}\\
A^{-+} & A^{--}
\end{array}\right),
\end{equation}
where 
\begin{equation}
\left(A^{\sigma_{1}\sigma_{2}}\right)_{k,l}^{\mu,\nu}=\frac{\partial\phi_{k}^{\sigma_{1},\mu}}{\partial u_{l}^{\sigma_{2},\nu}},
\end{equation}
where $\sigma_{1},\sigma_{2}\in\left\{ +,-\right\} $. The Gaudin
determinants are
\begin{equation}
\det G^{\pm}=\det(A^{++}\pm A^{+-}).
\end{equation}
It is crucial that we first take the derivative and only after take
the pair structure limit. 

\section{Extra selection rules}

Besides the pair structure, there may be other selection rules for
non-vanishing overlaps. These depend on the Hilbert space. In the
Hilbert space there is a special vector, the pseudo-vacuum $|0\rangle$
(Bethe vector without Bethe roots $|0\rangle\equiv|\emptyset\rangle$)
and we need the $\mathfrak{g}$ quantum numbers of the pseudo-vacuum.
These can be obtained from the series expansion of $\alpha$-functions
\begin{equation}
\alpha_{s}(u)=1+\frac{i}{u}d_{s}+\mathcal{O}(u^{-2}),
\end{equation}
for $s=1,\dots,n$. The $d_{s}$ are the Dynkin labels of the $\mathfrak{g}$-representation
corresponding to the pseudo-vacuum. We already showed a compatibility
condition for the Hilbert space, namely $\alpha_{s}(u)\alpha_{\rho(s)}(-u)=1$.
For chiral nodes it does not give any restriction for the Dynkin label
$d_{s}$ but it does for the achiral ones. For the achiral $\mathfrak{gl}_{N}$
MPS the Hilbert space has to satisfy $d_{N-s}=d_{s}$ where $s=1,\dots,N$
and for the achiral $\mathfrak{o}_{2n}$ MPS the Hilbert space has
to satisfy $d_{n-1}=d_{n}$.

In the following, we go through the possible reflection algebras and
the corresponding extra selection rules of the on-shell overlaps. 

\subsection*{$Y(\mathfrak{gl}_{N},\mathfrak{o}_{N})$ (chiral)}

For the $Y(\mathfrak{gl}_{N},\mathfrak{o}_{N})$ there is no extra
selection rule, therefore the pseudo-vacuum overlap $\mathbf{B}=\langle\Psi|0\rangle$
is non-vanishing.

\subsection*{$Y(\mathfrak{gl}_{2k},\mathfrak{sp}_{2k})$ (chiral)}

For the $Y(\mathfrak{gl}_{2k},\mathfrak{sp}_{2k})$ the overlap is
non-vanishing only when the numbers $r_{2a}$ are even for $a=1,\dots,k-1$,
and
\begin{equation}
r_{2a-1}=\frac{d_{2a-1}+r_{2a-2}+r_{2a}}{2},
\end{equation}
for $a=1,\dots,k$. We can see that the pseudo-vacuum overlap $\mathbf{B}=\langle\Psi|0\rangle$
is non-vanishing only when $d_{2a-1}=0$ for $a=1,\dots,k$.

\subsection*{$Y(\mathfrak{gl}_{N},\mathfrak{gl}_{M}\oplus\mathfrak{gl}_{N-M})$
(achiral)}

For the $Y(\mathfrak{gl}_{2n},\mathfrak{gl}_{n}\oplus\mathfrak{gl}_{n})$
or $Y(\mathfrak{gl}_{2n+1},\mathfrak{gl}_{n}\oplus\mathfrak{gl}_{n+1})$
the overlaps do not have extra selection rules. For $M<\left\lfloor \frac{N}{2}\right\rfloor $
the overlap is non-vanishing only when 
\begin{equation}
r_{j}=r_{M}+\sum_{k=M+1}^{j}\Lambda_{k},
\end{equation}
for $j=M+1,\dots,n_{+}$, where we introduced the weights
\begin{equation}
\Lambda_{j}=\begin{cases}
\sum_{k=j}^{n-1}d_{k}+\frac{1}{2}d_{n}, & N=2n,\\
\sum_{k=j}^{n}d_{k}, & N=2n+1,
\end{cases}
\end{equation}
for $j=M+1,\dots,n_{+}$. We can see that the pseudo-vacuum overlap
$\mathbf{B}=\langle\Psi|0\rangle$ is non-vanishing only when $d_{s}=0$
for $s=M+1,\dots,N-M$.

\subsection*{$Y(\mathfrak{o}_{2n+1},\mathfrak{o}_{M}\oplus\mathfrak{o}_{2n+1-M})$
(chiral)}

For$Y(\mathfrak{o}_{2n+1},\mathfrak{o}_{n}\oplus\mathfrak{o}_{n+1})$
the overlaps do not have extra selection rules. For $M<n$ the overlap
is non-vanishing only when 
\begin{equation}
r_{j}=r_{M}+\sum_{k=M+1}^{j}\Lambda_{k},
\end{equation}
for $j=M+1,\dots,n$, where we introduced the weights
\begin{equation}
\Lambda_{j}=\sum_{k=j}^{n}d_{k},
\end{equation}
for $j=M+1,\dots,n$. We can see that the pseudo-vacuum overlap $\mathbf{B}=\langle\Psi|0\rangle$
is non-vanishing only when $d_{s}=0$ for $s=M+1,\dots,n$.

\subsection*{$Y(\mathfrak{sp}_{2n},\mathfrak{gl}_{n})$ (chiral)}

For the $Y(\mathfrak{sp}_{2n},\mathfrak{gl}_{n})$ there is no extra
selection rule, therefore the pseudo-vacuum overlap $\mathbf{B}=\langle\Psi|0\rangle$
is non-vanishing.

\subsection*{$Y(\mathfrak{sp}_{2n},\mathfrak{sp}_{2m}\oplus\mathfrak{sp}_{2n-2m})$
(chiral)}

For$Y(\mathfrak{sp}_{2n},\mathfrak{sp}_{2m}\oplus\mathfrak{sp}_{2n-2m})$
the overlap is non-vanishing only when 
\begin{equation}
\begin{split}r_{2a-1} & =\frac{r_{2a-2}+r_{2a}+d_{2a-1}}{2},\quad\text{for }a=1,\dots,m,\\
r_{j} & =r_{2m}+\sum_{k=2m+1}^{j}\Lambda_{k},\quad j=2m+1,\dots,n-1,\\
r_{n} & =\frac{1}{2}r_{2m}+\frac{1}{2}\sum_{k=2m+1}^{n}\Lambda_{k},
\end{split}
\end{equation}
 where we introduced the weights
\begin{equation}
\Lambda_{j}=\sum_{k=j}^{n-1}d_{k}+\frac{1}{2}d_{n},
\end{equation}
for $j=M+1,\dots,n$. We can see that the pseudo-vacuum overlap $\mathbf{B}=\langle\Psi|0\rangle$
is non-vanishing only when $d_{2a-1}=d_{s}=0$ for $a=1,\dots,m$
and $s=2m+1,\dots,n$.

\subsection*{$Y(\mathfrak{o}_{2n},\mathfrak{o}_{M}\oplus\mathfrak{o}_{2n-M})$}

For$Y(\mathfrak{o}_{2n},\mathfrak{o}_{n}\oplus\mathfrak{o}_{n})$
(chiral) and $Y(\mathfrak{o}_{2n},\mathfrak{o}_{n-1}\oplus\mathfrak{o}_{n+1})$
(achiral) the overlaps do not have extra selection rules. For $M<n-1$
the overlap is non-vanishing only when 
\begin{equation}
\begin{split}r_{j} & =r_{M}+\sum_{k=M+1}^{j}\Lambda_{k},\quad j=M+1,\dots,n-2,\\
r_{n-1} & =\frac{1}{2}r_{M}+\frac{1}{2}\sum_{k=M+1}^{n-1}\Lambda_{k}-\frac{1}{2}\Lambda_{n},\\
r_{n} & =\frac{1}{2}r_{M}+\frac{1}{2}\sum_{k=M+1}^{n-1}\Lambda_{k}+\frac{1}{2}\Lambda_{n},
\end{split}
\end{equation}
where we introduced the weights
\begin{equation}
\begin{split}\Lambda_{j} & =\sum_{k=j}^{n-2}d_{k}+\frac{1}{2}d_{n-1}+\frac{1}{2}d_{n},\quad j=M+1,\dots,n-1,\\
\Lambda_{n} & =\frac{1}{2}d_{n}-\frac{1}{2}d_{n-1}.
\end{split}
\end{equation}
We can see that the pseudo-vacuum overlap $\mathbf{B}=\langle\Psi|0\rangle$
is non-vanishing only when $d_{s}=0$ for $s=M+1,\dots,n$.

\subsection*{$Y(\mathfrak{o}_{2n},\mathfrak{gl}_{n})$}

If $n$ is even (chiral) then
\begin{equation}
r_{2a-1}=\frac{r_{2a-2}+r_{2a}+d_{2a-1}}{2},
\end{equation}
for $a=1,\dots,\frac{n}{2}.$ When $n$ is odd (achiral) the selection
rules are
\begin{equation}
\begin{split}r_{2a-1} & =\frac{r_{2a-2}+r_{2a}+d_{2a-1}}{2},\quad\text{for }a=1,\dots,\frac{n-3}{2},\\
r_{n-2} & =\frac{r_{n-3}+2r_{n-1}+d_{n-2}}{2}.
\end{split}
\end{equation}
We can see that the pseudo-vacuum overlap $\mathbf{B}=\langle\Psi|0\rangle$
is non-vanishing only when $d_{2a-1}=0$ for $a=1,\dots,\frac{n}{2}$
when $n$ is even and for $a=1,\dots,\frac{n-1}{2}$ when $n$ is
odd.

\subsection*{Comments on the $\beta$-parameters of the MPS overlap function}

We just saw that for certain Hilbert spaces, the extra selection rules
lead to zero pseudo-vacuum overlap, i.e., $\mathbf{B}=\langle\Psi|0\rangle=0$.
Our general overlap formula (\ref{eq:ov}) is still valid, but for
these Hilbert spaces the $\beta$-parameters are not eigenvalues of
the operator $\mathbf{B}$. In these cases, the closed formula for
the $\beta$-parameters is not known (yet). Since these are simple
numbers (do not depending on the Bethe roots), their determination
is not too complicated in the concrete cases. They can be determined
by a simple numerical analysis.

To summarize, for Hilbert spaces where the pseudo-vacuum overlap $\mathbf{B}=\langle\Psi|0\rangle$
is non-zero, the $\beta$-parameters are eigenvalues of $\mathbf{B}$.
In the cases where the operator B is zero the determination of $\beta_{k}$-s
is a separate problem.

\section{$F$-operators for $\mathfrak{g}=\mathfrak{gl}_{N}$}

In the $\mathfrak{g}=\mathfrak{gl}_{N}$ case there exist chiral or
achiral pair structures. 

For chiral pair structure we introduce $N$ commuting operators $\mathbf{G}^{(k)}(u)$,
for $k=1,\dots,N$. We also introduce $F$-operators
\begin{equation}
\mathbf{F}^{(k)}(u)=\left[\mathbf{G}^{(k)}(u)\right]^{-1}\mathbf{G}^{(k+1)}(u),\label{eq:Fop}
\end{equation}
for $k=1,\dots,N-1$. They can be diagonalized simultaneously 
\begin{equation}
\mathbf{F}^{(k)}(u)=\mathbf{A}\mathrm{diag}(\mathcal{F}_{1}^{(k)}(u),\dots,\mathcal{F}_{d_{B}}^{(k)}(u))\mathbf{A}^{-1},\label{eq:Feig}
\end{equation}
where $\mathbf{A}$ is an invertible matrix of the boundary space.
The final $\tilde{\mathcal{F}}$ functions are defined as
\begin{equation}
\tilde{\mathcal{F}}_{j}^{(k)}(u)=\mathcal{F}_{j}^{(k)}(iu-k/2)\sqrt{\frac{u^{2}}{u^{2}+1/4}}.
\end{equation}
For achiral pair structure we introduce $n_{+}+1=\left\lfloor \frac{N}{2}\right\rfloor +1$
commuting operators $\mathbf{G}^{(k)}(u)$, for $k=1,\dots,n_{+}+1$.
The $F$-operators and their spectrum are defined in the same way
as before (\ref{eq:Fop}), (\ref{eq:Feig}) for $k=1,\dots,n_{+}$.
The final $\tilde{\mathcal{F}}$ functions defined as
\begin{equation}
\tilde{\mathcal{F}}_{j}^{(k)}(u)=\begin{cases}
\mathcal{F}_{j}^{(k)}(iu+\frac{N-2k}{4}), & k<\frac{N-1}{2},\\
\mathcal{F}_{j}^{(\frac{N-1}{2})}(iu+\frac{1}{4})\sqrt{\frac{u-i/4}{u+i/4}}, & k=\frac{N-1}{2},\\
\mathcal{F}_{j}^{(\frac{N}{2})}(iu)\sqrt{\frac{u^{2}}{u^{2}+1/4}}, & k=\frac{N}{2}.
\end{cases}
\end{equation}
The pseudo-vacuum overlap $\mathbf{B}=\langle\Psi|0\rangle$ also
commutes with the $F$-operators and its spectrum gives the $\beta_{j}$-s
of the overlap formula
\begin{equation}
\mathbf{B}=\mathbf{A}\mathrm{diag}(\beta_{1},\dots,\beta_{d_{B}})\mathbf{A}^{-1}.
\end{equation}

\subsection{$Y(\mathfrak{gl}_{N},\mathfrak{o}_{N})$ (chiral)\label{subsec:YpN}}
\begin{flushleft}
For the twisted Yangian $Y^{+}(N)\equiv Y(\mathfrak{gl}_{N},\mathfrak{o}_{N})$
we define a series of nested K-matrices with the following recursion
\begin{equation}
\mathbf{K}_{a,b}^{(k+1)}(u)=\mathbf{K}_{a,b}^{(k)}(u)-\mathbf{K}_{a,k}^{(k)}(u)\left[\mathbf{K}_{k,k}^{(k)}(u)\right]^{-1}\mathbf{K}_{k,b}^{(k)}(u),
\end{equation}
where $a,b=k+1,\dots,N$. The starting condition is $\mathbf{K}^{(1)}(u)\equiv\mathbf{K}(u)$.
We define the commuting operators as
\begin{equation}
\mathbf{G}^{(k)}(u)=\mathbf{K}_{k,k}^{(k)}(u),
\end{equation}
for $k=1,\dots,N$.
\par\end{flushleft}

\subsection{$Y(\mathfrak{gl}_{2k},\mathfrak{sp}_{2k})$ (chiral)\label{subsec:YmN}}

\subsubsection{$k=1$}

We start with the $k=1$ case, i.e. $Y^{-}(2)\equiv Y(\mathfrak{gl}_{2},\mathfrak{sp}_{2})$.
At first we have to find the factorization form
\begin{equation}
\mathbf{K}_{a,b}(u)=\sum_{c,d=1}^{2}\mathbf{L}_{c,a}(u)\epsilon_{c,d}\mathbf{L}_{d,b}(-u).
\end{equation}
We can define the deformed K-matrix
\begin{equation}
\tilde{\mathbf{K}}_{a,b}(u)=\mathbf{K}_{a,b}(u)+(u+1/2)\mathbf{L}_{1,a}(u)\mathbf{L}_{1,b}(-u).
\end{equation}
Now we can define the G-operators
\begin{equation}
\begin{split}\mathbf{G}^{(1)}(u) & =\tilde{\mathbf{K}}_{1,1}(u),\\
\mathbf{G}^{(2)}(u) & =\tilde{\mathbf{K}}_{2,2}(u)-\tilde{\mathbf{K}}_{2,1}(u)\left[\tilde{\mathbf{K}}_{1,1}(u)\right]^{-1}\tilde{\mathbf{K}}_{1,2}(u).
\end{split}
\end{equation}

\subsubsection{General $k$}
\begin{flushleft}
For general $k$, i.e. $Y^{-}(2k)\equiv Y(\mathfrak{gl}_{2k},\mathfrak{sp}_{2k})$,
the nesting is defined as
\begin{equation}
\mathbf{K}_{a,b}^{(s+1)}(u)=\mathbf{K}_{a,b}^{(s)}(u)-\sum_{\alpha,\beta=2s-1}^{2s}\mathbf{K}_{a,\alpha}^{(s)}(u)\widehat{\mathbf{K}}_{\alpha,\beta}^{(s)}(u)\mathbf{K}_{\beta,b}^{(s)}(u),
\end{equation}
for $a,b=2s+1,\dots,2k$. We defined the inverse matrix $\widehat{\mathbf{K}}_{\alpha,\beta}^{(s)}(u)$
for $\alpha,\beta=2s-1,2s$ as
\begin{equation}
\sum_{\beta=2s-1}^{2s}\widehat{\mathbf{K}}_{\alpha,\beta}^{(s)}(u)\mathbf{K}_{\beta,\gamma}^{(s)}(u)=\delta_{\alpha,\gamma}.
\end{equation}
We can select $k$ commuting $Y^{-}(2)$ subalgebras
\begin{equation}
\mathbf{k}^{(s)}(u)=\left(\begin{array}{cc}
\mathbf{K}_{2s-1,2s-1}^{(s)}(u) & \mathbf{K}_{2s-1,2s}^{(s)}(u)\\
\mathbf{K}_{2s,2s-1}^{(s)}(u) & \mathbf{K}_{2s,2s}^{(s)}(u)
\end{array}\right),
\end{equation}
where $s=1,\dots,k$. We can repeat the $Y^{-}(2)$ method separately.
We have to find the factorization form
\begin{equation}
\mathbf{k}_{a,b}^{(s)}(u)=\sum_{c,d=1}^{2}\mathbf{L}_{c,a}(u+s-1)\epsilon_{c,d}\mathbf{L}_{d,b}(-u-s+1),
\end{equation}
and the deformed K-matrix
\begin{equation}
\tilde{\mathbf{k}}_{a,b}^{(s)}(u)=\mathbf{k}_{a,b}^{(s)}(u)+(u+s-1/2)\mathbf{L}_{1,a}(u+s-1)\mathbf{L}_{1,b}(-u-s+1).
\end{equation}
Now we can define the G-operators
\begin{equation}
\begin{split}\mathbf{G}^{(2s-1)}(u) & =\tilde{\mathbf{k}}_{1,1}^{(s)}(u),\\
\mathbf{G}^{(2s)}(u) & =\tilde{\mathbf{k}}_{2,2}^{(s)}(u)-\tilde{\mathbf{k}}_{2,1}^{(s)}(u)\left[\tilde{\mathbf{k}}_{1,1}^{(s)}(u)\right]^{-1}\tilde{\mathbf{k}}_{1,2}^{(s)}(u),
\end{split}
\label{eq:GopYm}
\end{equation}
for $s=1,\dots,k$.
\par\end{flushleft}

\subsection{$Y(\mathfrak{gl}_{N},\mathfrak{gl}_{M}\oplus\mathfrak{gl}_{N-M})$
(achiral)\label{subsec:BNM}}

In this subsection, we use a symmetric convention for the indices
$a,b$ of the the K-matrix $\mathbf{K}_{a,b}$. For odd $N=2n+1$
the indices $a,b$ run through the set $\left\{ -n,...,-1,0,1,...,n\right\} $,
i.e. the K-matrix has the following components
\begin{equation}
\mathbf{K}=\left(\begin{array}{ccccccc}
\mathbf{K}_{-n,-n} & \dots & \mathbf{K}_{-n,-1} & \mathbf{K}_{-n,0} & \mathbf{K}_{-n,1} & \dots & \mathbf{K}_{-n,n}\\
\vdots & \ddots & \vdots & \vdots & \vdots & \iddots & \vdots\\
\mathbf{K}_{-1,-n} & \dots & \mathbf{K}_{-1,-1} & \mathbf{K}_{-1,0} & \mathbf{K}_{-1,1} & \dots & \mathbf{K}_{-1,n}\\
\mathbf{K}_{0,-n} & \dots & \mathbf{K}_{0,-1} & \mathbf{K}_{0,0} & \mathbf{K}_{0,1} & \dots & \mathbf{K}_{0,n}\\
\mathbf{K}_{1,-n} & \dots & \mathbf{K}_{1,-1} & \mathbf{K}_{1,0} & \mathbf{K}_{1,1} & \dots & \mathbf{K}_{1,n}\\
\vdots & \iddots & \vdots & \vdots & \vdots & \ddots & \vdots\\
\mathbf{K}_{n,-n} & \dots & \mathbf{K}_{n,-1} & \mathbf{K}_{n,0} & \mathbf{K}_{n,1} & \dots & \mathbf{K}_{n,n}
\end{array}\right).\label{eq:Kodd}
\end{equation}
For even $N=2n$ the indices $a,b$ run through the set $\left\{ -n,...,-1,1,...,n\right\} $,
i.e. the K-matrix has the following components
\begin{equation}
\mathbf{K}=\left(\begin{array}{cccccc}
\mathbf{K}_{-n,-n} & \dots & \mathbf{K}_{-n,-1} & \mathbf{K}_{-n,1} & \dots & \mathbf{K}_{-n,n}\\
\vdots & \ddots & \vdots & \vdots & \iddots & \vdots\\
\mathbf{K}_{-1,-n} & \dots & \mathbf{K}_{-1,-1} & \mathbf{K}_{-1,1} & \dots & \mathbf{K}_{-1,n}\\
\mathbf{K}_{1,-n} & \dots & \mathbf{K}_{1,-1} & \mathbf{K}_{1,1} & \dots & \mathbf{K}_{1,n}\\
\vdots & \iddots & \vdots & \vdots & \ddots & \vdots\\
\mathbf{K}_{n,-n} & \dots & \mathbf{K}_{n,-1} & \mathbf{K}_{n,1} & \dots & \mathbf{K}_{n,n}
\end{array}\right).\label{eq:Keven}
\end{equation}

\subsubsection{$Y(\mathfrak{gl}_{2n},\mathfrak{gl}_{n}\oplus\mathfrak{gl}_{n})$
or $Y(\mathfrak{gl}_{2n+1},\mathfrak{gl}_{n}\oplus\mathfrak{gl}_{n+1})$\label{subsec:BN}}
\begin{flushleft}
When $M=\left\lfloor \frac{N}{2}\right\rfloor $ we have a series
of $K$-matrices $\mathbf{K}^{(s)}$ for $s=1,\dots,n$ with the recursion
\begin{equation}
\mathbf{K}_{a,b}^{(s+1)}(u)=\mathbf{K}_{a,b}^{(s)}(u)-\mathbf{K}_{a,-n+s-1}^{(s)}(u)\left[\mathbf{K}_{n-s+1,-n+s-1}^{(s)}(u)\right]^{-1}\mathbf{K}_{n-s+1,b}^{(s)}(u),\label{eq:Breq}
\end{equation}
where $a,b=-n+s,-n+s+1,\dots,n-s-1,n-s$, with the initial condition
$\mathbf{K}^{(1)}(u)\equiv\mathbf{K}(u)$. We also define the G-operators
$\mathbf{G}^{(s)}$ for $s=1,\dots,n$ with 
\begin{equation}
\mathbf{G}^{(s)}(u)=\mathbf{K}_{n-s+1,-n+s-1}^{(s)}(u),\label{eq:Gdef}
\end{equation}
and 
\begin{equation}
\mathbf{G}^{(n+1)}(u)=\begin{cases}
\mathbf{K}_{0,0}^{(n)}(u)-\mathbf{K}_{0,-1}^{(n)}(u)\left[\mathbf{K}_{1,-1}^{(n)}(u)\right]^{-1}\mathbf{K}_{1,0}^{(n)}(u), & \text{for }N=2n+1,\\
\mathbf{K}_{-1,1}^{(n)}(u)-\mathbf{K}_{-1,-1}^{(n)}(u)\left[\mathbf{K}_{1,-1}^{(n)}(u)\right]^{-1}\mathbf{K}_{1,1}^{(n)}(u), & \text{for }N=2n.
\end{cases}\label{eq:Gnp}
\end{equation}
\par\end{flushleft}

\subsubsection{$M=0$\label{subsec:M0-1}}

For $M=0$. at first we have to find the factorization form
\begin{equation}
\mathbf{K}(u)=\mathbf{L}^{-1}(u)\mathbf{L}(-u).
\end{equation}
We can define the deformed K-matrix
\begin{equation}
\tilde{\mathbf{K}}_{a,b}(u)=\mathbf{K}_{a,b}(u)+u\sum_{c=1}^{n}\left[\mathbf{L}^{-1}(u)\right]_{a,c}\mathbf{L}_{-c,b}(-u).
\end{equation}
Now we can use the nesting (\ref{eq:Breq}) with the initial condition
$\mathbf{K}^{(1)}(u)\equiv\tilde{\mathbf{K}}(u)$ to obtain the nested
K-matrices $\mathbf{K}^{(s)}$ for $s=1,\dots,n$. We define the $G$-operators
with (\ref{eq:Gdef}), (\ref{eq:Gnp}).

\subsubsection{General $M$}
\begin{flushleft}
For general $M$ we combine the methods of sections \ref{subsec:BN}
and \ref{subsec:M0-1}. We have a series of $K$-matrices $\mathbf{K}^{(s)}$
for $s=1,\dots,M+1$ with the recursion (\ref{eq:Breq}) and initial
condition $\mathbf{K}^{(1)}(u)\equiv\mathbf{K}(u)$. We also define
the G-operators $\mathbf{G}^{(s)}$ for $s=1,\dots,M$ with (\ref{eq:Gdef}). 
\par\end{flushleft}

The K-matrix $\mathbf{K}^{(M+1)}$ is a $Y(\mathfrak{gl}_{N-2M},\mathfrak{gl}_{N-2M})$
representation therefore we continue with the method of \ref{subsec:M0-1}.
We have to find the factorization form
\begin{equation}
\mathbf{K}^{(M+1)}(u)=\mathbf{L}^{-1}(u)\mathbf{L}(-u),
\end{equation}
and deform the K-matrix
\begin{equation}
\tilde{\mathbf{K}}_{a,b}^{(M+1)}(u)=\mathbf{K}_{a,b}^{(M+1)}(u)+u\sum_{c=1}^{n-M}\left[\mathbf{L}^{-1}(u)\right]_{a,c}\mathbf{L}_{-c,b}(-u).
\end{equation}
We can continue the series of $K$-matrices $\mathbf{K}^{(s)}$ for
$s=M+2,\dots,n$ with the recursion (\ref{eq:Breq}) and initial condition
$\mathbf{K}^{(M+1)}(u)\equiv\tilde{\mathbf{K}}^{(M+1)}$. We define
the remaining $G$-operators as $\mathbf{G}^{(M+1)}(u)=\tilde{\mathbf{K}}_{n-M,-n+M}^{(M+1)}(u)$
and with the equation (\ref{eq:Gdef}) for $s=M+2,M+3,\dots,n$ and
the $\mathbf{G}^{(n+1)}(u)$ with (\ref{eq:Gnp}).

\section{$F$-operators for $\mathfrak{g}=\mathfrak{o}_{2n}$}

For $\mathfrak{g}=\mathfrak{o}_{2n}$ there exists chiral or achiral
pair structure. For the chiral case, we introduce $n+1$ commuting
operators $\mathbf{G}^{(k)}(u)$, for $k=1,\dots,n+1$. We also introduce
$F$-operators
\begin{equation}
\mathbf{F}^{(k)}(u)=\begin{cases}
\left[\mathbf{G}^{(k)}(u)\right]^{-1}\mathbf{G}^{(k+1)}(u), & k=1,\dots,n-1,\\
\left[\mathbf{G}^{(n-1)}(u)\right]^{-1}\mathbf{G}^{(n+1)}(u), & k=n,
\end{cases}\label{eq:Fop-oN}
\end{equation}
and the final $\tilde{\mathcal{F}}$ functions as
\begin{equation}
\tilde{\mathcal{F}}_{j}^{(k)}(u)=\begin{cases}
\mathcal{F}_{j}^{(k)}(iu+\frac{n-k-1}{2})\sqrt{\frac{u^{2}}{u^{2}+1/4}}, & k=1,\dots,n-2,\\
\mathcal{F}_{j}^{(k)}(iu)\sqrt{\frac{u^{2}}{u^{2}+1/4}}, & k=n-1,n.
\end{cases}
\end{equation}

For achiral pair structure we introduce $n$ commuting operators $\mathbf{G}^{(k)}(u)$
for $k=1,\dots,n$. The $F$-operators are defined as (\ref{eq:Fop})
for $k=1,\dots,n-1$. The final $\tilde{\mathcal{F}}$ functions defined
as
\begin{equation}
\tilde{\mathcal{F}}_{j}^{(k)}(u)=\begin{cases}
\mathcal{F}_{j}^{(k)}(iu+\frac{n-k-1}{2})\sqrt{\frac{u^{2}}{u^{2}+1/4}}, & k=1,\dots,n-2,\\
\mathcal{F}_{j}^{(n-1)}(iu)\frac{u}{u+i/2}, & k=n-1.
\end{cases}
\end{equation}

In this section, we use the convention of section \ref{subsec:BNM}
for the indices $a,b$ of the K-matrix $\mathbf{K}_{a,b}$, i.e.,
the indices $a,b$ run through the set $\left\{ -n,...,-1,1,...,n\right\} $,
and the K-matrix has the components (\ref{eq:Keven}).

\subsection{$Y(\mathfrak{o}_{2n},\mathfrak{gl}_{n})$}

\subsubsection{$Y(\mathfrak{o}_{4k},\mathfrak{gl}_{2k})$ (chiral)}

We define the nested $K$-matrices $\mathbf{K}^{(s)}$ for $s=2,\dots,k$
($\mathbf{K}^{(1)}\equiv\mathbf{K}$) with the recursion 
\begin{equation}
\mathbf{K}_{a,b}^{(s+1)}(u)=\mathbf{K}_{a,b}^{(s)}(u)-\sum_{\alpha,\beta=n-2s+1}^{n-2s+2}\mathbf{K}_{a,-\alpha}^{(s)}(u)\widehat{\mathbf{K}}_{-\alpha,\beta}^{(s)}(u)\mathbf{K}_{\beta,b}^{(s)}(u),\label{eq:KOrec}
\end{equation}
where $a,b=-n+2s,-n+2s+1,\dots,n-2s-1,n-2s$. We defined the inverse
matrix $\widehat{\mathbf{K}}_{-\alpha,\beta}^{(s)}(u)$ for $\alpha,\beta=n-2s+1,n-2s+2$
as
\begin{equation}
\sum_{\beta=n-2s+1}^{n-2s+2}\widehat{\mathbf{K}}_{-\alpha,\beta}^{(s)}(u)\mathbf{K}_{\beta,-\gamma}^{(s)}(u)=\delta_{\alpha,\gamma},
\end{equation}
where $s=1,\dots,k$. We can select commuting $Y^{-}(2)$ subalgebras
\begin{equation}
\mathbf{k}^{(s)}(u)=\left(\begin{array}{cc}
\mathbf{K}_{n-2s+2,-n+2s-2}^{(s)}(u) & \mathbf{K}_{n-2s+2,-n+2s-1}^{(s)}(u)\\
\mathbf{K}_{n-2s+1,-n+2s-2}^{(s)}(u) & \mathbf{K}_{n-2s+1,-n+2s-1}^{(s)}(u)
\end{array}\right),
\end{equation}
for $s=1,\dots,k$. We can use the $Y^{-}(2)$ method. We have to
find the deformed $K$-matrices
\begin{equation}
\begin{split}\mathbf{k}_{a,b}^{(s)}(u) & =\sum_{c,d=1}^{2}\mathbf{L}_{c,a}(u-\kappa_{N-4s+4})\epsilon_{c,d}\mathbf{L}_{d,b}(-u),\\
\tilde{\mathbf{k}}_{a,b}^{(s)}(u) & =\mathbf{k}_{a,b}^{(s)}(u)+(u+1/2-\kappa_{N-4s+4}/2)\mathbf{L}_{1,a}(u-\kappa_{N-4s+4})\mathbf{L}_{1,b}(-u),
\end{split}
\label{eq:kO}
\end{equation}
for $s=1,\dots,k$. The G-operators $\mathbf{G}^{(1)},\dots,\mathbf{G}^{(n)}$
are defined with (\ref{eq:GopYm}) for $s=1,\dots,k$. They define
the F-operators in the usual way
\begin{equation}
\mathbf{F}^{(s)}(u)=\left[\mathbf{G}^{(s)}(u)\right]^{-1}\mathbf{G}^{(s+1)}(u),\label{eq:Fop-1}
\end{equation}
for $s=1,\dots,n-1$. The remaining $F$-operator is
\begin{equation}
\mathbf{F}^{(n)}(u)=\left[\tilde{\mathbf{K}}_{2,-1}^{(k)}(u)\right]^{-1}\left[\tilde{\mathbf{K}}_{-1,2}^{(k)}(u)-\tilde{\mathbf{K}}_{-1,-1}^{(k)}(u)\left[\tilde{\mathbf{K}}_{2,-1}^{(k)}(u)\right]^{-1}\tilde{\mathbf{K}}_{2,2}^{(k)}(u)\right].\label{eq:Fopn}
\end{equation}

\subsubsection{$Y(\mathfrak{o}_{4k+2},\mathfrak{gl}_{2k+1})$ (achiral)}

We define the nested $K$-matrices $\mathbf{K}^{(s)}$ for $s=1,\dots,k$
with the recursion (\ref{eq:KOrec}) and initial condition $\mathbf{K}^{(1)}\equiv\mathbf{K}$.
We also define the $Y^{-}(2)$ $K$-matrices $\mathbf{k}^{(s)}$ the
deformed ones $\tilde{\mathbf{k}}^{(s)}$ for $s=1,\dots,k$ with
(\ref{eq:kO}). The G-operators $\mathbf{G}^{(1)},\dots,\mathbf{G}^{(n-1)}$
are defined with (\ref{eq:GopYm}) for $s=1,\dots,k$. The remaining
$G$-operator is
\begin{equation}
\mathbf{G}^{(n)}(u)=\mathbf{K}_{1,1}^{(k)}(u)-\sum_{\alpha,\beta=2}^{3}\mathbf{K}_{1,-\alpha}^{(k)}(u)\widehat{\mathbf{K}}_{-\alpha,\beta}^{(k)}(u)\mathbf{K}_{\beta,1}^{(k)}(u).\label{eq:Gnde}
\end{equation}

\subsection{$Y(\mathfrak{o}_{2n},\mathfrak{o}_{M}\oplus\mathfrak{o}_{2n-M})$}

\subsubsection{$Y(\mathfrak{o}_{2n},\mathfrak{o}_{n}\oplus\mathfrak{o}_{n})$ (chiral)}

We have a series of $K$-matrices $\mathbf{K}^{(s)}$ for $s=1,\dots,n-1$
with the recursion (\ref{eq:Breq}). We also define the G-operators
$\mathbf{G}^{(s)}$ for $s=1,\dots,n-1$ with (\ref{eq:Gdef}). We
also need to define the remaining $G$-operators $\mathbf{G}^{(n)}(u),\mathbf{G}^{(n+1)}(u)$
as
\begin{equation}
\begin{split}\mathbf{G}^{(n)}(u) & =\mathbf{K}_{1,-1}^{(n-1)}(u)-\mathbf{K}_{1,-2}^{(n-1)}(u)\left[\mathbf{K}_{2,-2}^{(n-1)}(u)\right]^{-1}\mathbf{K}_{2,-1}^{(n-1)}(u),\\
\mathbf{G}^{(n+1)}(u) & =\mathbf{K}_{-1,1}^{(n-1)}(u)-\mathbf{K}_{-1,-2}^{(n-1)}(u)\left[\mathbf{K}_{2,-2}^{(n-1)}(u)\right]^{-1}\mathbf{K}_{2,1}^{(n-1)}(u).
\end{split}
\end{equation}

\subsubsection{$Y(\mathfrak{o}_{2n},\mathfrak{o}_{n-1}\oplus\mathfrak{o}_{n+1})$
(achiral)}

We have a series of $K$-matrices $\mathbf{K}^{(s)}$ for $s=1,\dots,n-1$
with the recursion (\ref{eq:Breq}). We also define the G-operators
$\mathbf{G}^{(s)}$ for $s=1,\dots,n-1$ with (\ref{eq:Gdef}). We
also need to define the remaining $G$-operator $\mathbf{G}^{(n)}(u)$
as
\begin{equation}
\mathbf{G}^{(n)}(u)=\mathbf{K}_{1,1}^{(n-1)}(u)-\mathbf{K}_{1,-2}^{(n-1)}(u)\left[\mathbf{K}_{2,-2}^{(n-1)}(u)\right]^{-1}\mathbf{K}_{2,1}^{(n-1)}(u).
\end{equation}

\subsubsection{General $M$}

We continue with general $n$ and $M$. We have a series of $K$-matrices
$\mathbf{K}^{(s)}$ for $s=1,\dots,M+1$ with the recursion (\ref{eq:Breq}).
We also define the G-operators $\mathbf{G}^{(s)}$ for $s=1,\dots,M$
with (\ref{eq:Gdef}). We also need to define the remaining $G$-operators
$\mathbf{G}^{(M+1)}(u),\mathbf{G}^{(M+2)}(u),\dots,\mathbf{G}^{(n+1)}(u)$
for the chiral and $\mathbf{G}^{(M+1)}(u),\mathbf{G}^{(M+2)}(u),\dots,\mathbf{G}^{(n)}(u)$
for achiral cases. When $n-M$ is even the pair structure is chiral
and the pair structure is odd when $n-M$ is odd.

\textbf{If $n-M$ is even}, i.e. $n-M=2k$ ($k\in\mathbb{Z}_{+}$)
we have to deform the K-matrix $\mathbf{K}^{(M+1)}$ as 
\begin{equation}
\begin{split}\mathbf{K}_{a,b}^{(M+1)}(u) & =\sum_{c=-\tilde{n}}^{\tilde{n}}\mathbf{L}_{-c,-a}(u-\kappa_{N-2M})\mathbf{L}_{c,b}(-u),\\
\tilde{\mathbf{K}}_{a,b}^{(1)}(u) & =\mathbf{K}_{a,b}^{(M+1)}(u)+\\
 & +u\sum_{l=1}^{k}\left(\mathbf{L}_{-(\tilde{n}-2l+1),-a}(u-\kappa_{N-2M})\mathbf{L}_{-\tilde{n}+2l-2,b}(-u)-\mathbf{L}_{-(\tilde{n}-2l+2),-a}(u-\kappa_{N-2M})\mathbf{L}_{-\tilde{n}+2l-1,b}(-u)\right),
\end{split}
\label{eq:KOdef}
\end{equation}
where $\tilde{n}=2k=n-M$. We continue with the series of the nested
$K$-matrices $\tilde{\mathbf{K}}_{a,b}^{(s)}(u)$ for $s=2,\dots,k$
as (\ref{eq:KOrec}) and define the $Y^{-}(2)$ $K$-matrices $\mathbf{k}^{(s)}$
and the deformed ones $\tilde{\mathbf{k}}^{(s)}$ for $s=1,\dots,k$
with (\ref{eq:kO}) (we replace $n\to\tilde{n}$ , $\mathbf{K}^{(s)}\to\tilde{\mathbf{K}}^{(s)}$
and $\kappa_{N-4s+4}\to\kappa_{N-2M-4s+4}$). The G-operators $\mathbf{G}^{(M+1)},\dots,\mathbf{G}^{(n)}$
are
\begin{equation}
\begin{split}\mathbf{G}^{(M+2s-1)}(u) & =\tilde{\mathbf{k}}_{1,1}^{(s)}(u),\\
\mathbf{G}^{(M+2s)}(u) & =\tilde{\mathbf{k}}_{2,2}^{(s)}(u)-\tilde{\mathbf{k}}_{2,1}^{(s)}(u)\left[\tilde{\mathbf{k}}_{1,1}^{(s)}(u)\right]^{-1}\tilde{\mathbf{k}}_{1,2}^{(s)}(u),
\end{split}
\label{eq:GopO-2-1}
\end{equation}
 for $s=1,\dots,k$. The F-operators are defined as (\ref{eq:Fop-1})
and (\ref{eq:Fopn}).

\textbf{If $n-M$ is odd}, i.e. $n-M=2k+1$ ($k\in\mathbb{Z}_{+}$)
we have to deform the K-matrix $\mathbf{K}^{(M+1)}$ as (\ref{eq:KOdef})
where $\tilde{n}=2k+1=n-M$. We continue with the series of the nested
$K$-matrices $\tilde{\mathbf{K}}_{a,b}^{(s)}(u)$ for $s=2,\dots,k$
as (\ref{eq:KOrec}). We define the $Y^{-}(2)$ $K$-matrices $\mathbf{k}^{(s)}$
and the deformed ones $\tilde{\mathbf{k}}^{(s)}$ for $s=1,\dots,k$
with (\ref{eq:kO}) (we replace $n\to\tilde{n}$ , $\mathbf{K}^{(s)}\to\tilde{\mathbf{K}}^{(s)}$
and $\kappa_{N-4s+4}\to\kappa_{N-2M-4s+4}$). The G-operators $\mathbf{G}^{(M+1)},\dots,\mathbf{G}^{(n-1)}$
are defined with (\ref{eq:GopO-2-1}) for $s=1,\dots,k$. The remaining
$G$-operator $\mathbf{G}^{(n)}$ is defined as (\ref{eq:Gnde}).

\section{$F$-operators for $\mathfrak{g}=\mathfrak{o}_{2n+1}$}

In the $\mathfrak{g}=\mathfrak{o}_{2n+1}$ case there exists only
chiral pair structure. We introduce $n+1$ $G$-operators $\mathbf{G}^{(k)}(u)$
for $k=1,\dots,n+1$ and we introduce the $F$-operators as (\ref{eq:Fop}).
The final $\tilde{\mathcal{F}}$ functions are defined as
\begin{equation}
\tilde{\mathcal{F}}_{j}^{(k)}(u)=\begin{cases}
\mathcal{F}_{j}^{(k)}(iu-\frac{1}{4}+\frac{n-k}{2})\sqrt{\frac{u^{2}}{u^{2}+1/4}}, & k=1,\dots,n-1,\\
\mathcal{F}_{j}^{(n)}(iu-\frac{1}{4})\frac{u+i/4}{u+i/2}\sqrt{\frac{u^{2}}{u^{2}+1/16}}, & k=n.
\end{cases}
\end{equation}

In this section, we use the convention of section \ref{subsec:BNM}
for the indices $a,b$ of the K-matrix $\mathbf{K}_{a,b}$, i.e.,
the indices $a,b$ run through the set $\left\{ -n,...,-1,0,1,...,n\right\} $,
and the K-matrix has the components (\ref{eq:Kodd}).

\subsection{$Y(\mathfrak{o}_{2n+1},\mathfrak{o}_{n}\oplus\mathfrak{o}_{n+1})$}

For $Y(\mathfrak{o}_{2n+1},\mathfrak{o}_{n}\oplus\mathfrak{o}_{n+1})$
we have a series of $K$-matrices $\mathbf{K}^{(k)}$ for $k=1,\dots,n$
with the recursion (\ref{eq:Breq}) and initial condition $\mathbf{K}^{(1)}(u)\equiv\mathbf{K}(u)$.
We also define the G-operators $\mathbf{G}^{(k)}$ for $k=1,\dots,n$
with (\ref{eq:Gdef}) and $\mathbf{G}^{(n+1)}$ with (\ref{eq:Gnp}). 

\subsection{$Y(\mathfrak{o}_{2n+1},\mathfrak{o}_{M}\oplus\mathfrak{o}_{2n-M+1})$}

For general $M$ we have a series of $K$-matrices $\mathbf{K}^{(s)}$
for $s=1,\dots,M+1$ with the recursion (\ref{eq:Breq}) and initial
condition $\mathbf{K}^{(1)}(u)\equiv\mathbf{K}(u)$. We also define
the G-operators $\mathbf{G}^{(s)}$ for $s=1,\dots,M$ with (\ref{eq:Gdef}).
We also need to define the remaining $G$-operators $\mathbf{G}^{(M+1)}(u),\mathbf{G}^{(M+2)}(u),\dots,\mathbf{G}^{(n+1)}(u)$.
Their definitions depend on whether $n-M$ is even or odd.

\textbf{If $n-M$ is odd}, i.e. $n-M=2k+1$ ($k\in\mathbb{N}$) we
have to deform the K-matrix $\mathbf{K}^{(M+1)}$ with (\ref{eq:KOdef})
where $\tilde{n}=2k+1=n-M$. We define the nested $K$-matrices $\tilde{\mathbf{K}}_{a,b}^{(s)}(u)$
for $s=2,\dots,k+1$ with (\ref{eq:KOrec}), the $Y^{-}(2)$ $K$-matrices
$\mathbf{k}^{(s)}$ and the deformed ones $\tilde{\mathbf{k}}^{(s)}$
for $s=1,\dots,k$ with (\ref{eq:kO}) (we replace $n\to\tilde{n}$
, $\mathbf{K}^{(s)}\to\tilde{\mathbf{K}}^{(s)}$ and $\kappa_{N-4s+4}\to\kappa_{N-2M-4s+4}$).
We can also define the G-operators $\mathbf{G}^{(r)}$ for $r=M+1,\dots,n-1$
as (\ref{eq:GopO-2-1}) for $s=1,\dots,k$. We also have to deform
the K-matrix $\tilde{\mathbf{K}}_{a,b}^{(k+1)}(u)$ as 
\begin{equation}
\begin{split}\tilde{\mathbf{K}}_{a,b}^{(k+1)}(u) & =\sum_{c=-1}^{1}\mathbf{L}_{-c,-a}(u-\kappa_{3})\mathbf{L}_{c,b}(-u),\\
\check{\mathbf{K}}_{a,b}^{(k+1)}(u) & =\tilde{\mathbf{K}}_{a,b}^{(k+1)}(u)-\frac{1}{2}u(u+\frac{1}{4})\mathbf{L}_{-1,-a}(u-\kappa_{3})\mathbf{L}_{-1,b}(-u)\\
 & +u\left(\mathbf{L}_{0,-a}(u-\kappa_{3})\mathbf{L}_{-1,b}(-u)-\mathbf{L}_{-1,-a}(u-\kappa_{3})\mathbf{L}_{0,b}(-u)\right).
\end{split}
\label{eq:KO4k1}
\end{equation}
 and the remaining $G$-operators $\mathbf{G}^{(n)}$, $\mathbf{G}^{(n+1)}$
are
\begin{equation}
\begin{split}\mathbf{G}^{(n)}(u) & =\check{\mathbf{K}}_{1,-1}^{(k+1)}(u),\\
\mathbf{G}^{(n+1)}(u) & =\check{\mathbf{K}}_{0,0}^{(k+1)}(u)-\check{\mathbf{K}}_{0,-1}^{(k+1)}(u)\left[\check{\mathbf{K}}_{1,-1}^{(k+1)}(u)\right]^{-1}\check{\mathbf{K}}_{1,0}^{(k+1)}(u).
\end{split}
\label{eq:G4k1}
\end{equation}

\textbf{If $n-M$ is even}, i.e. $n-M=2k$ ($k\in\mathbb{Z}_{+}$),
we have to deform the K-matrix $\mathbf{K}^{(M+1)}$ with (\ref{eq:KOdef})
where we replace $n\to\tilde{n}=2k=n-M$ , $\mathbf{K}\to\mathbf{K}^{(M+1)}$
and $\kappa_{N}\to\kappa_{N-2M}$. We also define the nested $K$-matrices
$\tilde{\mathbf{K}}^{(s)}$ for $s=1,\dots,k$ with the recursion
(\ref{eq:KOrec}) and the $Y^{-}(2)$ $K$-matrices $\mathbf{k}^{(s)}$
the deformed ones $\tilde{\mathbf{k}}^{(s)}$ for $s=1,\dots,k$ with
(\ref{eq:kO}). The G-operators $\mathbf{G}^{(M+1)},\dots,\mathbf{G}^{(n)}$
are defined with (\ref{eq:GopO-2-1}) for $s=1,\dots,k$. The remaining
$G$-operator $\mathbf{G}^{(n+1)}$ is defined as 
\begin{equation}
\mathbf{G}^{(n+1)}(u)=\mathbf{K}_{0,0}^{(k)}(u)-\sum_{\alpha,\beta=1}^{2}\mathbf{K}_{0,-\alpha}^{(k)}(u)\widehat{\mathbf{K}}_{-\alpha,\beta}^{(k)}(u)\mathbf{K}_{\beta,0}^{(k)}(u).\label{eq:Gn1}
\end{equation}

\section{$F$-operators for $\mathfrak{g}=\mathfrak{sp}_{2n}$}

In the $\mathfrak{g}=\mathfrak{sp}_{2n}$ case there exists only chiral
pair structure. We introduce $n+1$ G-operators $\mathbf{G}^{(k)}(u)$
for $k=1,\dots,n+1$ and we introduce the $F$-operators as (\ref{eq:Fop})
and the final $\tilde{\mathcal{F}}$ functions are defined as
\begin{equation}
\tilde{\mathcal{F}}_{j}^{(k)}(u)=\begin{cases}
\mathcal{F}_{j}^{(k)}(iu+\frac{n-k}{2})\sqrt{\frac{u^{2}}{u^{2}+1/4}}, & k=1,\dots,n-1,\\
\mathcal{F}_{j}^{(n)}(iu)\sqrt{\frac{u^{2}}{u^{2}+1}}, & k=n.
\end{cases}
\end{equation}
In this section, we use the convention of section \ref{subsec:BNM}
for the indices $a,b$ of the K-matrix $\mathbf{K}_{a,b}$, i.e.,
the indices $a,b$ run through the set $\left\{ -n,...,-1,1,...,n\right\} $,
and the K-matrix has the components (\ref{eq:Keven}).

\subsection{$Y(\mathfrak{sp}_{2n},\mathfrak{gl}_{n})$}

We have a series of $K$-matrices $\mathbf{K}^{(k)}$ for $k=1,\dots,n$
with the recursion (\ref{eq:Breq}) and initial condition $\mathbf{K}^{(1)}(u)\equiv\mathbf{K}(u)$.
We also define the G-operators $\mathbf{G}^{(k)}$ for $k=1,\dots,n$
with (\ref{eq:Gdef}) and $\mathbf{G}^{(n+1)}$ with (\ref{eq:Gnp}). 

\subsection{$Y(\mathfrak{sp}_{2n},\mathfrak{sp}_{2m}\oplus\mathfrak{sp}_{2n-2m})$}

We have a series of $K$-matrices $\mathbf{K}^{(s)}$ for $s=1,\dots,m+1$
with the recursion (\ref{eq:KOrec}) and initial condition $\mathbf{K}^{(1)}\equiv\mathbf{K}$.
We also define the $Y^{-}(2)$ $K$-matrices $\mathbf{k}^{(s)}$ and
the deformed ones $\tilde{\mathbf{k}}^{(s)}$ for $s=1,\dots,m$ with
(\ref{eq:kO}). We can define the G-operators $\mathbf{G}^{(1)},\dots,\mathbf{G}^{(2m)}$
with (\ref{eq:GopYm}) for $s=1,\dots,m$. For $n>2m$ we also need
to define the remaining $G$-operators $\mathbf{G}^{(2m+1)}(u),\mathbf{G}^{(2m+2)}(u),\dots,\mathbf{G}^{(n+1)}(u)$.

For $k=n-2m>0$, we have to deform the K-matrix $\mathbf{K}^{(2m+1)}$
as
\begin{equation}
\begin{split}\mathbf{K}_{a,b}^{(2m+1)}(u) & =\sum_{c=-k}^{k}\theta_{a}\theta_{c}\mathbf{L}_{-c,-a}(u-\kappa_{2k})\mathbf{L}_{c,b}(-u),\\
\tilde{\mathbf{K}}_{a,b}^{(1)}(u) & =\mathbf{K}_{a,b}^{(2m+1)}(u)+\sum_{c=1}^{k}\theta_{a}\theta_{c}\mathbf{L}_{-c,-a}(u-\kappa_{2k})\mathbf{L}_{-c,b}(-u),
\end{split}
\end{equation}
where
\begin{equation}
\theta_{i}=\begin{cases}
+1, & i>0,\\
-1, & i<0.
\end{cases}
\end{equation}
We also define the nested $K$-matrices $\tilde{\mathbf{K}}^{(s)}$
for $s=2,\dots,k$ with the recursion (\ref{eq:Breq}). The G-operators
$\mathbf{G}^{(2m+1)},\dots,\mathbf{G}^{(n)}$ are defined as
\begin{equation}
\mathbf{G}^{(2m+s)}(u)=\tilde{\mathbf{K}}_{k-s+1,-k+s-1}^{(s)}(u),
\end{equation}
for $s=1,\dots,k$ and
\begin{equation}
\mathbf{G}^{(n+1)}(u)=\tilde{\mathbf{K}}_{-1,1}^{(k)}(u)-\tilde{\mathbf{K}}_{-1,-1}^{(k)}(u)\left[\tilde{\mathbf{K}}_{1,-1}^{(k)}(u)\right]^{-1}\tilde{\mathbf{K}}_{1,1}^{(k)}(u).
\end{equation}

\section{Examples}

In this section, we demonstrate the application of the overlap formulas
on concrete examples. 

\subsection{$SO(6)$ spin chain}

Let us consider the $SO(6)$ spin chain which is given by the Hamiltonian
\begin{equation}
H=\sum_{j=1}^{J}\mathbf{P}_{j,j+1}-\frac{1}{2}\mathbf{Q}_{j,j+1},
\end{equation}
where 
\begin{equation}
\mathbf{P}=\sum_{i,j=1}^{6}e_{i,j}\otimes e_{j,i},\quad\mathbf{Q}=\sum_{i,j=1}^{6}e_{i,j}\otimes e_{7-i,7-j}.
\end{equation}
In the AdS/CFT context we often identify the states as the scalar
fields of the $\mathcal{N}=4$ SYM as 
\begin{equation}
|1\rangle\equiv Z,\ensuremath{\quad}|2\rangle\equiv Y,\ensuremath{\quad}|3\rangle\equiv X,\ensuremath{\quad}|4\rangle\equiv\bar{X},\ensuremath{\quad}|5\rangle\equiv\bar{Y},\ensuremath{\quad}|6\rangle\equiv\bar{Z}.
\end{equation}
 The zeroth step to apply our formula is to get monodromy matrices.
Since the $\mathfrak{sl}_{4}\cong\mathfrak{o}_{6}$ we could use the
$\mathfrak{gl}_{4}$ or $\mathfrak{o}_{6}$ formulas. Now we continue
with the $\mathfrak{gl}_{4}$ convention. The $\mathfrak{gl}_{4}$
monodromy matrices can be defined in the following way. Let $e_{i}$
be the usual basis in $\mathbb{C}^{4}$. Let $e_{i,j}$ be the unit
matrices in $\mathbb{C}^{4}$ as $e_{i,j}e_{k}=\delta_{j,k}e_{i}$.
Let $e_{[i,j]}$ for $1\leq i<j\leq4$ be a basis of $\mathbb{C}^{6}$
which coming from the anti-symmetrization of $\mathbb{C}^{4}\otimes\mathbb{C}^{4}$.
Let $e_{[i,j],[k,l]}$ be the unit matrices in $\mathbb{C}^{6}$ as
$e_{[i,j],[k,l]}e_{[a,b]}=\delta_{k,a}\delta_{l,b}e_{[i,j]}$. We
can identify the six dimensional basis as
\begin{align}
Z & \equiv e_{[1,2]}, & \bar{X} & \equiv e_{[2,3]},\nonumber \\
Y & \equiv e_{[1,3]}, & \bar{Y} & \equiv e_{[4,2]},\\
X & \equiv e_{[1,4]}, & \bar{Z} & \equiv e_{[3,4]}.\nonumber 
\end{align}
Let us define the 6-dimensional matrices $\mathcal{E}_{i,j}$ as
\begin{equation}
\mathcal{E}_{i,j}=\sum_{c=1}^{4}e_{[i,c],[j,c]}.
\end{equation}
Let us define the following Lax-operators
\begin{equation}
\mathcal{L}(u)=\sum_{i,j=1}^{4}e_{i,j}\otimes\left(\delta_{i,j}+\frac{1}{u+1/2}\mathcal{E}_{j,i}\right),\qquad\widehat{\mathcal{L}}(u)=\sum_{i,j=1}^{4}e_{i,j}\otimes\left(\delta_{i,j}+\frac{1}{-u+1/2}\mathcal{E}_{i,j}\right).
\end{equation}
The monodromy matrices
\begin{align}
T_{0}(u) & =\mathcal{L}_{0,J}(u)\dots\mathcal{L}_{0,2}(u)\mathcal{L}_{0,1}(u),\nonumber \\
\widehat{T}_{0}(u) & =\widehat{\mathcal{L}}_{0,J}(u)\dots\widehat{\mathcal{L}}_{0,2}(u)\widehat{\mathcal{L}}_{0,1}(u).
\end{align}
satisfy the $\mathfrak{gl}_{4}$ $RTT$-relations.

\subsubsection{$\mathfrak{o}_{4}=\mathfrak{o}_{3}\oplus\mathfrak{o}_{3}$ symmetric
boundary states}

Our MPS is built from a one-side operator
\begin{equation}
\begin{split}\bar{\omega} & =\left\{ S_{1},S_{2},S_{3},S_{3},S_{2},S_{1}\right\} ,\\
\langle\mathrm{MPS}| & =\sum_{i_{1},\dots,i_{J}}\mathrm{Tr}\left[\omega_{i_{J}}\dots\omega_{i_{1}}\right]\langle i_{1},i_{2},\dots,i_{J-1},i_{J}|.
\end{split}
\end{equation}
We defined the spin operators $S_{1},S_{2},S_{3}$ which satisfy the
$\mathfrak{su}_{2}$ algebra
\begin{equation}
\left[S_{j},S_{k}\right]=\sum_{l=1}^{3}i\epsilon_{j,k,l}S_{l}.
\end{equation}
We will also need the $\mathfrak{su}_{2}$ Casimir operator
\begin{equation}
S^{2}=S_{1}^{2}+S_{2}^{2}+S_{3}^{2}.
\end{equation}
We have finite dimensional MPS for every integer of half integer $s$
for which Casimir is $s(s+1)$.

The first step is to find the $K$-matrix. The $K$-matrix is the
solution of the $KT$-relation for $J=1$
\begin{equation}
\sum_{k=1}^{4}\sum_{a=1}^{6}\mathbf{K}_{i,k}(u)\omega_{a}\mathcal{L}_{k,j;a,b}(u)=\sum_{k=1}^{4}\sum_{a=1}^{6}\omega_{a}\widehat{\mathcal{L}}_{i,k;a,b}(-u)\mathbf{K}_{k,j}(u).
\end{equation}
The solution is
\begin{equation}
\mathbf{K}(u)=\left(\begin{array}{cccc}
1 & \frac{2i}{2u+1}S_{1} & \frac{2i}{2u+1}S_{2} & \frac{2i}{2u+1}S_{3}\\
\frac{-2i}{2u+1}S_{1} & 1 & \frac{2i}{2u+1}S_{3} & \frac{-2i}{2u+1}S_{2}\\
\frac{-2i}{2u+1}S_{2} & \frac{-2i}{2u+1}S_{3} & 1 & \frac{2i}{2u+1}S_{1}\\
\frac{-2i}{2u+1}S_{3} & \frac{2i}{2u+1}S_{2} & \frac{-2i}{2u+1}S_{1} & 1
\end{array}\right).
\end{equation}

Next step is to calculate the $G$-operators which are defined in
\ref{subsec:YpN}. We need the nested $K$-matrices $\mathbf{K}^{(2)},\mathbf{K}^{(3)}$.
Since $\mathbf{K}_{1,1}=1$ the second $K$-matrix is simply defined
as
\begin{equation}
\mathbf{K}_{a,b}^{(2)}(u)=\mathbf{K}_{a,b}(u)-\mathbf{K}_{a,1}(u)\mathbf{K}_{1,b}(u),
\end{equation}
for $a,b=2,3,4$ i.e. the $3\times3$ matrix is 
\begin{equation}
\mathbf{K}^{(2)}=\left(\begin{array}{ccc}
\mathbf{K}_{2,2}^{(2)} & \mathbf{K}_{2,3}^{(2)} & \mathbf{K}_{2,4}^{(2)}\\
\mathbf{K}_{3,2}^{(2)} & \mathbf{K}_{3,3}^{(2)} & \mathbf{K}_{3,4}^{(2)}\\
\mathbf{K}_{4,2}^{(2)} & \mathbf{K}_{4,3}^{(2)} & \mathbf{K}_{4,4}^{(2)}
\end{array}\right)=\left(\begin{array}{ccc}
1-\frac{4S_{1}^{2}}{(2u+1)^{2}} & \frac{2i}{2u+1}S_{3}-\frac{4S_{1}S_{2}}{(2u+1)^{2}} & \frac{-2i}{2u+1}S_{2}-\frac{4S_{1}S_{3}}{(2u+1)^{2}}\\
\frac{-2i}{2u+1}S_{3}-\frac{4S_{2}S_{1}}{(2u+1)^{2}} & 1-\frac{4S_{2}^{2}}{(2u+1)^{2}} & \frac{2i}{2u+1}S_{1}-\frac{4S_{2}S_{3}}{(2u+1)^{2}}\\
\frac{2i}{2u+1}S_{2}-\frac{4S_{3}S_{1}}{(2u+1)^{2}} & \frac{-2i}{2u+1}S_{1}-\frac{4S_{3}S_{2}}{(2u+1)^{2}} & 1-\frac{4S_{3}^{2}}{(2u+1)^{2}}
\end{array}\right).
\end{equation}
The third $K$-matrix 
\begin{equation}
\mathbf{K}_{a,b}^{(3)}(u)=\mathbf{K}_{a,b}^{(2)}(u)-\mathbf{K}_{a,2}^{(2)}(u)\left[\mathbf{K}_{2,2}^{(2)}(u)\right]^{-1}\mathbf{K}_{2,b}^{(2)}(u)
\end{equation}
is more problematic because of the inverse. We can handle it with
the identity
\begin{equation}
\mathbf{K}_{2,2}^{(2)}(u+1)\mathbf{K}_{a,2}^{(2)}(u)=\left(\frac{2u+3}{2u+2}\mathbf{K}_{a,2}^{(2)}(u+1)-\frac{1}{2u+2}\mathbf{K}_{2,a}^{(2)}(u+1)\right)\mathbf{K}_{2,2}^{(2)}(u),
\end{equation}
therefore
\begin{equation}
\mathbf{K}_{a,2}^{(2)}(u)\left[\mathbf{K}_{2,2}^{(2)}(u)\right]^{-1}=\left[\mathbf{K}_{2,2}^{(2)}(u+1)\right]^{-1}\left(\frac{2u+3}{2u+2}\mathbf{K}_{a,2}^{(2)}(u+1)-\frac{1}{2u+2}\mathbf{K}_{2,a}^{(2)}(u+1)\right).
\end{equation}
Substituting back, the third K-matrix is simplified as
\begin{equation}
\mathbf{K}_{a,b}^{(3)}(u)=\left[\mathbf{K}_{2,2}^{(2)}(u+1)\right]^{-1}\tilde{\mathbf{K}}_{a,b}^{(3)}(u),
\end{equation}
where we introduced a $2\times2$ matrix without inversion:
\begin{equation}
\tilde{\mathbf{K}}_{a,b}^{(3)}(u)=\mathbf{K}_{2,2}^{(2)}(u+1)\mathbf{K}_{a,b}^{(2)}(u)-\frac{2u+3}{2u+2}\mathbf{K}_{a,2}^{(2)}(u+1)\mathbf{K}_{2,b}^{(2)}(u)+\frac{1}{2u+2}\mathbf{K}_{2,a}^{(2)}(u+1)\mathbf{K}_{2,b}^{(2)}(u).
\end{equation}
After substitution 
\begin{equation}
\tilde{\mathbf{K}}^{(3)}=\left(\begin{array}{cc}
\tilde{\mathbf{K}}_{3,3}^{(3)} & \tilde{\mathbf{K}}_{3,4}^{(3)}\\
\tilde{\mathbf{K}}_{4,3}^{(3)} & \tilde{\mathbf{K}}_{4,4}^{(3)}
\end{array}\right)=\frac{(2u+1)(2u+3)-4S^{2}}{(2u+1)(2u+3)}\left(\begin{array}{cc}
1 & \frac{2i}{2u+3}S_{1}\\
\frac{-2i}{2u+3}S_{1} & 1
\end{array}\right).
\end{equation}
Now we can calculate the $G$-operators
\begin{equation}
\begin{split}\mathbf{G}^{(1)}(u)= & \mathbf{K}_{1,1}(u)=1,\\
\mathbf{G}^{(2)}(u)= & \mathbf{K}_{2,2}^{(2)}(u)=\frac{(2u+1)^{2}-4S_{1}^{2}}{(2u+1)^{2}},\\
\mathbf{G}^{(3)}(u)= & \mathbf{K}_{3,3}^{(3)}(u)=\frac{(2u+3)}{(2u+1)}\frac{(2u+1)(2u+3)-4S^{2}}{(2u+3)^{2}-4S_{1}^{2}},\\
\mathbf{G}^{(4)}(u)= & \mathbf{K}_{4,4}^{(4)}(u)=\mathbf{K}_{4,4}^{(3)}(u)-\mathbf{K}_{4,3}^{(3)}(u)\left[\mathbf{K}_{3,3}^{(3)}(u)\right]^{-1}\mathbf{K}_{3,4}^{(3)}(u)\\
= & \frac{(2u+1)(2u+3)-4S^{2}}{(2u+3)(2u+1)}.
\end{split}
\end{equation}
The F-operators are
\begin{equation}
\begin{split}\mathbf{F}^{(1)}(u)= & \left[\mathbf{G}^{(1)}(u)\right]^{-1}\mathbf{G}^{(2)}(u)=\frac{(2u+1)^{2}-4S_{1}^{2}}{(2u+1)^{2}},\\
\mathbf{F}^{(2)}(u)= & \left[\mathbf{G}^{(2)}(u)\right]^{-1}\mathbf{G}^{(3)}(u)=\frac{(2u+1)(2u+3)}{(2u+1)^{2}-4S_{1}^{2}}\frac{(2u+1)(2u+3)-4S^{2}}{(2u+3)^{2}-4S_{1}^{2}}\\
\mathbf{F}^{(3)}(u)= & \left[\mathbf{G}^{(3)}(u)\right]^{-1}\mathbf{G}^{(4)}(u)=\frac{(2u+3)^{2}-4S_{1}^{2}}{(2u+3)^{2}},
\end{split}
,
\end{equation}
and the pseudo-vacuum ($|0\rangle=Z^{J}$) overlap is
\begin{equation}
\mathbf{B}=S_{1}^{J}.
\end{equation}
The spectrum of the operators $S_{1},S^{2}$ can be labeled by $m=-s,-s+1,\dots,s-1,s$,
therefore the eigenvalues are 
\begin{equation}
\begin{split}\mathcal{F}_{m}^{(1)}(u) & =\frac{(2u+1-2m)(2u+1+2m)}{(2u+1)^{2}},\\
\mathcal{F}_{m}^{(2)}(u)= & \frac{(2u+1)(2u+3)}{(2u+1-2m)(2u+1+2m)}\frac{(2u+1-2s)(2u+3+2s)}{(2u+3-2m)(2u+3+2m)},\\
\mathcal{F}_{m}^{(3)}(u)= & \frac{(2u+3-2m)(2u+3+2m)}{(2u+3)^{2}}.
\end{split}
\end{equation}
therefore
\begin{equation}
\begin{split}\beta_{m} & =m^{J},\\
\tilde{\mathcal{F}}_{m}^{(1)}(u) & =\frac{u^{2}+m^{2}}{u^{2}}\sqrt{\frac{u^{2}}{u^{2}+1/4}},\\
\mathcal{\tilde{\mathcal{F}}}_{m}^{(2)}(u) & =\frac{(u^{2}+1/4)(u^{2}+(s+1/2)^{2})}{(u^{2}+(m+1/2)^{2})(u^{2}+(m-1/2)^{2})}\sqrt{\frac{u^{2}}{u^{2}+1/4}},\\
\mathcal{\tilde{\mathcal{F}}}_{m}^{(3)}(u) & =\frac{u^{2}+m^{2}}{u^{2}}\sqrt{\frac{u^{2}}{u^{2}+1/4}}.
\end{split}
\end{equation}

\subsubsection{$\mathfrak{sp}_{4}=\mathfrak{o}_{5}$ symmetric boundary states}

Our MPS is built from a one-side operator
\begin{equation}
\begin{split}\bar{\omega} & =\left\{ \gamma_{5},\gamma_{1}+i\gamma_{4},\gamma_{2}+i\gamma_{3},\gamma_{2}-i\gamma_{3},\gamma_{1}-i\gamma_{4},\gamma_{5}\right\} \\
\langle\mathrm{MPS}| & =\sum_{i_{1},\dots,i_{J}}\mathrm{Tr}\left[\omega_{i_{J}}\dots\omega_{i_{1}}\right]\langle i_{1},i_{2},\dots,i_{J-1},i_{J}|.
\end{split}
\end{equation}
We defined the $4\times4$ $\gamma$-matrices which satisfy the algebra
\begin{equation}
\left\{ \gamma_{j},\gamma_{k}\right\} =2\delta_{j,k}.
\end{equation}
We can use the concrete representation
\begin{equation}
\begin{split}\gamma_{1} & =\left(\begin{array}{cccc}
0 & 0 & 1 & 0\\
0 & 0 & 0 & 1\\
1 & 0 & 0 & 0\\
0 & 1 & 0 & 0
\end{array}\right),\;\gamma_{2}=\left(\begin{array}{cccc}
0 & 0 & 0 & i\\
0 & 0 & i & 0\\
0 & -i & 0 & 0\\
-i & 0 & 0 & 0
\end{array}\right),\;\gamma_{3}=\left(\begin{array}{cccc}
0 & 0 & 0 & 1\\
0 & 0 & -1 & 0\\
0 & -1 & 0 & 0\\
1 & 0 & 0 & 0
\end{array}\right),\\
\gamma_{4} & =\left(\begin{array}{cccc}
0 & 0 & i & 0\\
0 & 0 & 0 & -i\\
-i & 0 & 0 & 0\\
0 & i & 0 & 0
\end{array}\right),\;\gamma_{5}=\left(\begin{array}{cccc}
1 & 0 & 0 & 0\\
0 & 1 & 0 & 0\\
0 & 0 & -1 & 0\\
0 & 0 & 0 & -1
\end{array}\right).
\end{split}
\end{equation}

The first step is to find the $K$-matrix. The $K$-matrix is the
solution of the $KT$-relation for $J=1$
\begin{equation}
\sum_{k=1}^{4}\sum_{a=1}^{6}\mathbf{K}_{i,k}(u)\omega_{a}\mathcal{L}_{k,j;a,b}(u)=\sum_{k=1}^{4}\sum_{a=1}^{6}\omega_{a}\widehat{\mathcal{L}}_{i,k;a,b}(-u)\mathbf{K}_{k,j}(u).
\end{equation}
The solution is
\begin{equation}
\mathbf{K}(u)=\left(\begin{array}{cccc}
\left(\begin{array}{cccc}
0 & 0 & 0 & 0\\
0 & 0 & 0 & 0\\
0 & 0 & 0 & a_{4}\\
0 & 0 & 0 & 0
\end{array}\right) & \left(\begin{array}{cccc}
1 & 0 & 0 & 0\\
0 & 1 & 0 & 0\\
0 & 0 & a_{3} & 0\\
0 & 0 & 0 & a_{2}
\end{array}\right) & \left(\begin{array}{cccc}
0 & 0 & 0 & 0\\
0 & 0 & 0 & a_{1}\\
-a_{5} & 0 & 0 & 0\\
0 & 0 & 0 & 0
\end{array}\right) & \left(\begin{array}{cccc}
0 & 0 & 0 & ia_{1}(u)\\
0 & 0 & 0 & 0\\
0 & ia_{5} & 0 & 0\\
0 & 0 & 0 & 0
\end{array}\right)\\
\left(\begin{array}{cccc}
-1 & 0 & 0 & 0\\
0 & -1 & 0 & 0\\
0 & 0 & -a_{2} & 0\\
0 & 0 & 0 & -a_{3}
\end{array}\right) & \left(\begin{array}{cccc}
0 & 0 & 0 & 0\\
0 & 0 & 0 & 0\\
0 & 0 & 0 & 0\\
0 & 0 & a_{4} & 0
\end{array}\right) & \left(\begin{array}{cccc}
0 & 0 & 0 & 0\\
0 & 0 & ia_{1} & 0\\
0 & 0 & 0 & 0\\
ia_{5} & 0 & 0 & 0
\end{array}\right) & \left(\begin{array}{cccc}
0 & 0 & -a_{1} & 0\\
0 & 0 & 0 & 0\\
0 & 0 & 0 & 0\\
0 & a_{5} & 0 & 0
\end{array}\right)\\
\left(\begin{array}{cccc}
0 & 0 & 0 & 0\\
0 & 0 & 0 & a_{5}\\
-a_{1} & 0 & 0 & 0\\
0 & 0 & 0 & 0
\end{array}\right) & \left(\begin{array}{cccc}
0 & 0 & 0 & 0\\
0 & 0 & ia_{5} & 0\\
0 & 0 & 0 & 0\\
ia_{1} & 0 & 0 & 0
\end{array}\right) & \left(\begin{array}{cccc}
0 & 0 & 0 & 0\\
a_{4} & 0 & 0 & 0\\
0 & 0 & 0 & 0\\
0 & 0 & 0 & 0
\end{array}\right) & \left(\begin{array}{cccc}
-a_{2} & 0 & 0 & 0\\
0 & -a_{3} & 0 & 0\\
0 & 0 & -1 & 0\\
0 & 0 & 0 & -1
\end{array}\right)\\
\left(\begin{array}{cccc}
0 & 0 & 0 & ia_{5}\\
0 & 0 & 0 & 0\\
0 & ia_{1} & 0 & 0\\
0 & 0 & 0 & 0
\end{array}\right) & \left(\begin{array}{cccc}
0 & 0 & -a_{5} & 0\\
0 & 0 & 0 & 0\\
0 & 0 & 0 & 0\\
0 & a_{1} & 0 & 0
\end{array}\right) & \left(\begin{array}{cccc}
a_{3} & 0 & 0 & 0\\
0 & a_{2} & 0 & 0\\
0 & 0 & 1 & 0\\
0 & 0 & 0 & 1
\end{array}\right) & \left(\begin{array}{cccc}
0 & a_{4} & 0 & 0\\
0 & 0 & 0 & 0\\
0 & 0 & 0 & 0\\
0 & 0 & 0 & 0
\end{array}\right)
\end{array}\right),
\end{equation}
where
\begin{align}
a_{1} & =\frac{1}{u+1}, & a_{4} & =\frac{-i(2u+1)}{(u-1)(u+1)},\nonumber \\
a_{2} & =\frac{u}{u+1}, & a_{5} & =\frac{(u+2)}{(u-1)(u+1)},\\
a_{3} & =\frac{u^{2}+u+1}{(u-1)(u+1)}.\nonumber 
\end{align}

Next step is to calculate the $G$-operators which are defined in
\ref{subsec:YmN}. The first $Y^{-}(2)$ $K$-matrix
\[
\mathbf{k}^{(1)}(u)=\left(\begin{array}{cc}
\mathbf{k}_{1,1}^{(1)}(u) & \mathbf{k}_{1,2}^{(1)}(u)\\
\mathbf{k}_{2,1}^{(1)}(u) & \mathbf{k}_{2,2}^{(1)}(u)
\end{array}\right)=\left(\begin{array}{cc}
\mathbf{K}_{1,1}^{(1)}(u) & \mathbf{K}_{1,2}^{(1)}(u)\\
\mathbf{K}_{2,1}^{(1)}(u) & \mathbf{K}_{2,2}^{(1)}(u)
\end{array}\right),
\]
can be expressed with the Lax operator

\begin{equation}
\mathbf{L}(u)=\left(\begin{array}{cc}
\mathbf{L}_{1,1}(u) & \mathbf{L}_{1,2}(u)\\
\mathbf{L}_{2,1}(u) & \mathbf{L}_{2,2}(u)
\end{array}\right)=\left(\begin{array}{cc}
\left(\begin{array}{cccc}
1 & 0 & 0 & 0\\
0 & 1 & 0 & 0\\
0 & 0 & \frac{u}{u-1} & 0\\
0 & 0 & 0 & 1
\end{array}\right) & \left(\begin{array}{cccc}
0 & 0 & 0 & 0\\
0 & 0 & 0 & 0\\
0 & 0 & 0 & 0\\
0 & 0 & \frac{-i}{u-1} & 0
\end{array}\right)\\
\left(\begin{array}{cccc}
0 & 0 & 0 & 0\\
0 & 0 & 0 & 0\\
0 & 0 & 0 & \frac{i}{u-1}\\
0 & 0 & 0 & 0
\end{array}\right) & \left(\begin{array}{cccc}
1 & 0 & 0 & 0\\
0 & 1 & 0 & 0\\
0 & 0 & 1 & 0\\
0 & 0 & 0 & \frac{u}{u-1}
\end{array}\right)
\end{array}\right),
\end{equation}
as
\begin{equation}
\mathbf{k}_{a,b}^{(1)}(u)=\sum_{c,d=1}^{2}\mathbf{L}_{c,a}(u)\epsilon_{c,d}\mathbf{L}_{d,b}(-u).
\end{equation}
We can define the deformed K-matrix as
\[
\tilde{\mathbf{k}}_{a,b}^{(1)}(u)=\mathbf{k}_{a,b}^{(1)}(u)+(u+1/2)\mathbf{L}_{1,a}(u)\mathbf{L}_{1,b}(-u).
\]
The first two $G$-operators are
\begin{equation}
\begin{split}\mathbf{G}^{(1)}(u) & =\tilde{\mathbf{k}}_{1,1}^{(1)}(u)=\left(\begin{array}{cccc}
u+1/2 & 0 & 0 & 0\\
0 & u+1/2 & 0 & 0\\
0 & 0 & \frac{u^{2}(u+1/2)}{(u-1)(u+1)} & \frac{-i(2u+1)}{(u-1)(u+1)}\\
0 & 0 & 0 & u+1/2
\end{array}\right),\\
\mathbf{G}^{(2)}(u) & =\tilde{\mathbf{k}}_{2,2}^{(1)}(u)-\tilde{\mathbf{k}}_{2,1}^{(1)}(u)\left[\tilde{\mathbf{k}}_{1,1}^{(1)}(u)\right]^{-1}\tilde{\mathbf{k}}_{1,2}^{(1)}(u)=\left(\begin{array}{cccc}
\frac{1}{u+1/2} & 0 & 0 & 0\\
0 & \frac{1}{u+1/2} & 0 & 0\\
0 & 0 & \frac{u(u+2)}{(u+1)^{2}(u+1/2)} & \frac{2i}{(u+1)^{2}(u+1/2)}\\
0 & 0 & 0 & \frac{1}{u+1/2}
\end{array}\right).
\end{split}
\label{eq:GopYm-1}
\end{equation}
 We also need the nested $K$-matrix $\mathbf{K}^{(2)}$
\begin{equation}
\mathbf{K}_{a,b}^{(2)}(u)=\mathbf{K}_{a,b}(u)-\sum_{\alpha,\beta=1}^{2}\mathbf{K}_{a,\alpha}(u)\widehat{\mathbf{K}}_{\alpha,\beta}^{(1)}(u)\mathbf{K}_{\beta,b}(u),
\end{equation}
for $a,b=3,4$. We need the inverse matrix
\begin{equation}
\widehat{\mathbf{K}}^{(1)}(u)=\left(\begin{array}{cc}
\widehat{\mathbf{K}}_{1,1}^{(1)}(u) & \widehat{\mathbf{K}}_{1,2}^{(1)}(u)\\
\widehat{\mathbf{K}}_{2,1}^{(1)}(u) & \widehat{\mathbf{K}}_{2,2}^{(1)}(u)
\end{array}\right)=\left(\begin{array}{cc}
\left(\begin{array}{cccc}
0 & 0 & 0 & 0\\
0 & 0 & 0 & 0\\
0 & 0 & 0 & 0\\
0 & 0 & \frac{-i(2u+1)}{u(u+2)} & 0
\end{array}\right) & \left(\begin{array}{cccc}
-1 & 0 & 0 & 0\\
0 & -1 & 0 & 0\\
0 & 0 & -\frac{u+1}{u} & 0\\
0 & 0 & 0 & -\frac{u^{2}+u+1}{u(u+2)}
\end{array}\right)\\
\left(\begin{array}{cccc}
1 & 0 & 0 & 0\\
0 & 1 & 0 & 0\\
0 & 0 & \frac{u^{2}+u+1}{u(u+2)} & 0\\
0 & 0 & 0 & \frac{u+1}{u}
\end{array}\right) & \left(\begin{array}{cccc}
0 & 0 & 0 & 0\\
0 & 0 & 0 & 0\\
0 & 0 & 0 & \frac{-i(2u+1)}{u(u+2)}\\
0 & 0 & 0 & 0
\end{array}\right)
\end{array}\right).
\end{equation}
After substitution we obtain the second $K$-matrix
\begin{multline}
\mathbf{K}^{(2)}(u)=\left(\begin{array}{cc}
\mathbf{K}_{3,3}^{(2)}(u) & \mathbf{K}_{3,4}^{(2)}(u)\\
\mathbf{K}_{4,3}^{(2)}(u) & \mathbf{K}_{4,4}^{(2)}(u)
\end{array}\right)=\left(\begin{array}{cc}
\mathbf{k}_{1,1}^{(2)}(u) & \mathbf{k}_{1,2}^{(2)}(u)\\
\mathbf{k}_{2,1}^{(2)}(u) & \mathbf{k}_{2,2}^{(2)}(u)
\end{array}\right)=\\
\left(\begin{array}{cc}
\left(\begin{array}{cccc}
0 & 0 & 0 & 0\\
\frac{-i(2u+3)}{(u+1)^{2}} & 0 & 0 & 0\\
0 & 0 & 0 & 0\\
0 & 0 & 0 & 0
\end{array}\right) & \left(\begin{array}{cccc}
-\frac{u}{u+1} & 0 & 0 & 0\\
0 & -\frac{u^{2}+3u+3}{(u+1)^{2}} & 0 & 0\\
0 & 0 & -\frac{u(u+2)}{(u+1)^{2}} & 0\\
0 & 0 & 0 & -\frac{u(u+2)}{(u+1)^{2}}
\end{array}\right)\\
\left(\begin{array}{cccc}
\frac{u^{2}+3u+3}{(u+1)^{2}} & 0 & 0 & 0\\
0 & \frac{u}{u+1} & 0 & 0\\
0 & 0 & \frac{u(u+2)}{(u+1)^{2}} & 0\\
0 & 0 & 0 & \frac{u(u+2)}{(u+1)^{2}}
\end{array}\right) & \left(\begin{array}{cccc}
0 & \frac{-i(2u+3)}{(u+1)^{2}} & 0 & 0\\
0 & 0 & 0 & 0\\
0 & 0 & 0 & 0\\
0 & 0 & 0 & 0
\end{array}\right)
\end{array}\right).
\end{multline}
With the Lax operator

\begin{equation}
\mathbf{L}(u)=\left(\begin{array}{cc}
\mathbf{L}_{1,1}(u) & \mathbf{L}_{1,2}(u)\\
\mathbf{L}_{2,1}(u) & \mathbf{L}_{2,2}(u)
\end{array}\right)=i\left(\begin{array}{cc}
\left(\begin{array}{cccc}
1 & 0 & 0 & 0\\
0 & \frac{u+1}{u} & 0 & 0\\
0 & 0 & \frac{u-1}{u} & 0\\
0 & 0 & 0 & \frac{u-1}{u}
\end{array}\right) & \left(\begin{array}{cccc}
0 & \frac{i}{u} & 0 & 0\\
0 & 0 & 0 & 0\\
0 & 0 & 0 & 0\\
0 & 0 & 0 & 0
\end{array}\right)\\
\left(\begin{array}{cccc}
0 & 0 & 0 & 0\\
-\frac{i}{u} & 0 & 0 & 0\\
0 & 0 & 0 & 0\\
0 & 0 & 0 & 0
\end{array}\right) & \left(\begin{array}{cccc}
\frac{u+1}{u} & 0 & 0 & 0\\
0 & 1 & 0 & 0\\
0 & 0 & \frac{u-1}{u} & 0\\
0 & 0 & 0 & \frac{u-1}{u}
\end{array}\right)
\end{array}\right),
\end{equation}
we obtain the factorized form and the deformed matrix:
\begin{equation}
\begin{split}\mathbf{k}_{a,b}^{(2)}(u) & =\sum_{c,d=1}^{2}\mathbf{L}_{c,a}(u+1)\epsilon_{c,d}\mathbf{L}_{d,b}(-u-1),\\
\tilde{\mathbf{k}}_{a,b}^{(2)}(u) & =\mathbf{k}_{a,b}^{(2)}(u)+(u+3/2)\mathbf{L}_{1,a}(u+1)\mathbf{L}_{1,b}(-u-1).
\end{split}
\end{equation}
The remaining $G$-operators are
\begin{equation}
\begin{split}\mathbf{G}^{(3)}(u) & =\tilde{\mathbf{k}}_{1,1}^{(2)}(u)=\left(\begin{array}{cccc}
-(u+3/2) & 0 & 0 & 0\\
\frac{-i(2u+3)}{(u+1)^{2}} & -\frac{u(u+2)(u+3/2)}{(u+1)^{2}} & 0 & 0\\
0 & 0 & -\frac{u(u+2)(u+3/2)}{(u+1)^{2}} & 0\\
0 & 0 & 0 & -\frac{u(u+2)(u+3/2)}{(u+1)^{2}}
\end{array}\right),\\
\mathbf{G}^{(4)}(u) & =\tilde{\mathbf{k}}_{2,2}^{(2)}(u)-\tilde{\mathbf{k}}_{2,1}^{(2)}(u)\left[\tilde{\mathbf{k}}_{1,1}^{(2)}(u)\right]^{-1}\tilde{\mathbf{k}}_{1,2}^{(2)}(u)=\\
 & =\left(\begin{array}{cccc}
-\frac{u(u+3)}{(u+1)(u+2)(u+3/2)} & 0 & 0 & 0\\
\frac{2iu}{(u+1)^{2}(u+2)(u+3/2)} & -\frac{u(u+2)}{(u+1)^{2}(u+3/2)} & 0 & 0\\
0 & 0 & -\frac{u(u+2)}{(u+1)^{2}(u+3/2)} & 0\\
0 & 0 & 0 & -\frac{u(u+2)}{(u+1)^{2}(u+3/2)}
\end{array}\right).
\end{split}
\label{eq:GopYm-1-1}
\end{equation}
Using the basis transformation
\begin{equation}
\mathbf{A}=\left(\begin{array}{cccc}
0 & 0 & 0 & 1\\
0 & 0 & 1 & 2i\\
1 & 2i & 0 & 0\\
0 & 1 & 0 & 0
\end{array}\right),
\end{equation}
the $B$- and $F$-operators have diagonal form and the spectrum is
\begin{equation}
\begin{split}\mathbf{B} & =\mathbf{A}\mathrm{diag}\left((-1)^{J},(-1)^{J},1,1\right)\mathbf{A}^{-1},\\
\mathbf{F}^{(1)}(iu-1/2) & =\mathbf{A}\mathrm{diag}\left(-\frac{1}{u^{2}}\frac{(u^{2}+9/4)}{(u^{2}+1/4)},-\frac{1}{u^{2}},-\frac{1}{u^{2}},-\frac{1}{u^{2}}\right)\mathbf{A}^{-1},\\
\mathbf{F}^{(2)}(iu-2/2) & =\mathbf{A}\mathrm{diag}\left((u^{2}+1/4),(u^{2}+1/4)\frac{(u^{2}+1)}{u^{2}},(u^{2}+1/4)\frac{(u^{2}+1)}{u^{2}},(u^{2}+1/4)\right)\mathbf{A}^{-1},\\
\mathbf{F}^{(3)}(iu-3/2) & =\mathbf{A}\mathrm{diag}\left(-\frac{1}{u^{2}},-\frac{1}{u^{2}},-\frac{1}{u^{2}},-\frac{1}{u^{2}}\frac{(u^{2}+9/4)}{(u^{2}+1/4)}\right)\mathbf{A}^{-1}.
\end{split}
\end{equation}

\subsection{Alternating $SU(4)$ spin chain}

The Hamiltonian of the alternating $SU(4)$ spin chain is the following
\begin{equation}
H=\sum_{j=1}^{2J}1-\mathbf{P}_{j,j+2}+\frac{1}{2}\mathbf{P}_{j,j+2}\mathbf{Q}_{j,j+1}+\frac{1}{2}\mathbf{Q}_{j,j+1}\mathbf{P}_{j,j+2},
\end{equation}
where 
\begin{equation}
\mathbf{P}=\sum_{i,j=1}^{4}e_{i,j}\otimes e_{j,i},\quad\mathbf{Q}=\sum_{i,j=1}^{4}e_{i,j}\otimes e_{i,j}.
\end{equation}
In the AdS/CFT context we often identify the states as the scalar
fields of the ABJM as $|a\rangle\equiv Y^{a}$ for the odd sites $|a\rangle\equiv Y_{a}^{\dagger}$
for the even sites. The zeroth step to apply our formula is to get
the monodromy matrix. The $\mathfrak{gl}_{4}$ monodromy matrix can
be defined in the following way. Let us define the following Lax-operators
\begin{equation}
\mathcal{L}(u)=\sum_{i,j=1}^{4}e_{i,j}\otimes\left(\delta_{i,j}+\frac{1}{u-1}e_{j,i}\right),\qquad\widehat{\mathcal{L}}(u)=\sum_{i,j=1}^{4}e_{i,j}\otimes\left(\delta_{i,j}+\frac{1}{-u-1}e_{i,j}\right).
\end{equation}
The monodromy matrix
\begin{equation}
T_{0}(u)=\widehat{\mathcal{L}}_{0,2J}(u)\mathcal{L}_{0,2J-1}(u)\dots\widehat{\mathcal{L}}_{0,2}(u)\mathcal{L}_{0,1}(u)
\end{equation}
satisfies the $\mathfrak{gl}_{4}$ $RTT$-relation.

\subsubsection{$\mathfrak{gl}_{2}\oplus\mathfrak{gl}_{2}$ symmetric boundary states}

Our MPS is built from a two-side operator
\begin{equation}
\begin{split}\psi_{a,b} & =\left(\begin{array}{cccc}
s+S_{3}+1 & S_{1}+iS_{2} & 0 & 0\\
S_{1}-iS_{2} & s-S_{3}+1 & 0 & 0\\
0 & 0 & 0 & 0\\
0 & 0 & 0 & 0
\end{array}\right)_{a,b},\\
\langle\mathrm{MPS}| & =\sum_{i_{1},\dots,i_{J}}\mathrm{Tr}\left[\psi_{i_{2J-1},i_{2J}}\dots\psi_{i_{1},i_{2}}\right]\langle i_{1},i_{2},\dots,i_{2J-1},i_{2J}|,
\end{split}
\end{equation}
where $S_{1},S_{2},S_{3}$ are the spin operators for the irreducible
spin $s$ representation.

The first step is to find the $K$-matrix. The $K$-matrix is the
solution of the $KT$-relation for $J=1$
\begin{equation}
\sum_{a,b,k,l=1}^{4}\mathbf{K}_{i,k}(u)\psi_{a,b}\widehat{\mathcal{L}}_{k,l;b,d}(u)\mathcal{L}_{l,j;a,c}(u)=\sum_{a,b,k,l=1}^{4}\psi_{a,b}\widehat{\mathcal{L}}_{i,k;b,d}(-u)\mathcal{L}_{k,l;a,c}(-u)\mathbf{K}_{l,j}(u).
\end{equation}
The solution is
\begin{equation}
\mathbf{K}(u)=\left(\begin{array}{cccc}
a(u)+uS_{3} & uS_{-} & 0 & 0\\
uS_{+} & a(u)-uS_{3} & 0 & 0\\
0 & 0 & b(u) & 0\\
0 & 0 & 0 & b(u)
\end{array}\right),
\end{equation}
where $S_{\pm}=S_{1}\pm iS_{2}$ and 
\begin{equation}
\begin{split}a(u) & =-\frac{1}{2}u^{2}+\frac{1}{2}u+s+1,\\
b(u) & =(u-1)(\frac{1}{2}u-s-1).
\end{split}
\end{equation}

Next step is to calculate the $G$-operators which are defined in
\ref{subsec:BNM}. We can see that the $\mathbf{K}_{4,1}$ component
is not invertible therefore we have to rotate the boundary state and
the $K$-matrix. We use the $GL(4)$ transformation 
\begin{equation}
g=\left(\begin{array}{cccc}
1 & 0 & 0 & 0\\
0 & 1 & 0 & 0\\
0 & -\frac{\mathfrak{b}_{2}}{2} & 1 & 0\\
-\frac{\mathfrak{b}_{1}}{2} & 0 & 0 & 1
\end{array}\right).
\end{equation}
 The co-product is
\begin{equation}
\Delta(g)=\left(g\otimes\left(g^{-1}\right)^{t}\right)^{\otimes J},
\end{equation}
for which 
\begin{equation}
\left[\Delta(g),H\right]=0.
\end{equation}
The $K$-matrix and the boundary state transforms as
\begin{equation}
\begin{split}\psi_{a,b}^{g} & =\sum_{c,d=1}^{4}\psi_{c,d}g_{c,a}\left(g^{-1}\right)_{b,d},\\
\mathbf{K}^{g}(u) & =\left(g^{-1}\otimes\mathbf{1}\right)\mathbf{K}(u)\left(g\otimes\mathbf{1}\right).
\end{split}
\end{equation}
In the following we work with the transformed quantities and take
the $\mathfrak{b}_{1},\mathfrak{b}_{2}\to0$ limit only for the final
formula. 

The pseudo-vacuum is $|0\rangle=\left(|1\rangle\otimes|4\rangle\right)^{\otimes J}$
and the pseudo-vacuum overlap is
\begin{equation}
\mathbf{B}=\langle\Psi^{g}|0\rangle=(\psi_{1,4}^{g})^{J}=\mathfrak{b}_{1}^{J}\left(\frac{S_{3}+s+1}{2}\right)^{J}.
\end{equation}
We need the nested $K$-matrix 
\begin{equation}
\mathbf{K}_{a,b}^{(2)}(u)=\mathbf{K}_{a,b}^{g}(u)-\mathbf{K}_{a,1}^{g}(u)\left[\mathbf{K}_{4,1}^{g}(u)\right]^{-1}\mathbf{K}_{4,b}^{g}(u),
\end{equation}
for $a,b=2,3$. We can handle the inverse with the identity
\begin{equation}
\mathbf{K}_{4,1}^{g}(u+1)\mathbf{K}_{a,1}^{g}(u)=\mathbf{K}_{a,1}^{g}(u+1)\mathbf{K}_{4,1}^{g}(u),
\end{equation}
for $a=2,3$, therefore
\begin{equation}
\mathbf{K}_{a,1}^{g}(u)\left[\mathbf{K}_{4,1}^{g}(u)\right]^{-1}=\left[\mathbf{K}_{4,1}^{g}(u+1)\right]^{-1}\mathbf{K}_{a,1}^{g}(u+1).
\end{equation}
Substituting back, the nested K-matrix is simplified as
\begin{equation}
\mathbf{K}_{a,b}^{(2)}(u)=\left[\mathbf{K}_{4,1}^{g}(u+1)\right]^{-1}\tilde{\mathbf{K}}_{a,b}^{(2)}(u),
\end{equation}
where we introduced a $2\times2$ matrix without inversion:
\begin{equation}
\tilde{\mathbf{K}}_{a,b}^{(2)}(u)=\mathbf{K}_{4,1}^{g}(u+1)\mathbf{K}_{a,b}^{g}(u)-\mathbf{K}_{a,1}^{g}(u+1)\mathbf{K}_{4,b}^{g}(u).
\end{equation}
After substitution 
\begin{equation}
\tilde{\mathbf{K}}^{(2)}=\left(\begin{array}{cc}
\tilde{\mathbf{K}}_{2,2}^{(2)} & \tilde{\mathbf{K}}_{2,3}^{(2)}\\
\tilde{\mathbf{K}}_{3,2}^{(2)} & \tilde{\mathbf{K}}_{3,3}^{(2)}
\end{array}\right)=\frac{\mathfrak{b}_{1}\mathfrak{b}_{2}}{4}u(u^{2}-1)(u-2s-2)\left(\begin{array}{cc}
\frac{1}{\mathfrak{b}_{2}}\frac{S_{3}+s+1}{u}+\frac{1}{\mathfrak{b}_{2}} & 0\\
1 & \frac{1}{\mathfrak{b}_{2}}\frac{S_{3}+s+1}{u}-\frac{1}{\mathfrak{b}_{2}}
\end{array}\right).
\end{equation}
Now we can calculate the $G$-operators
\begin{equation}
\begin{split}\mathbf{G}^{(1)}(u)= & \mathbf{K}_{4,1}^{g}(u)=-\frac{\mathfrak{b}_{1}}{2}u(u-S_{3}-s-2),\\
\mathbf{G}^{(2)}(u)= & \mathbf{K}_{3,2}^{(2)}(u)=-\frac{\mathfrak{b}_{2}}{2}\frac{u(u-1)(u-2s-2)}{(u-S_{3}-s-1)},\\
\mathbf{G}^{(3)}(u)= & \mathbf{K}_{2,3}^{(2)}(u)-\mathbf{K}_{2,2}^{(2)}(u)\left[\mathbf{K}_{3,2}^{(2)}(u)\right]^{-1}\mathbf{K}_{3,3}^{(2)}(u)=\\
=- & \frac{1}{2\mathfrak{b}_{2}}\frac{(u-1)(u-2s-2)(u+S_{3}+s+1)}{u}.
\end{split}
\end{equation}
The F-operators are
\begin{equation}
\begin{split}\mathbf{F}^{(1)}(u)= & \left[\mathbf{G}^{(1)}(u)\right]^{-1}\mathbf{G}^{(2)}(u)=\frac{\mathfrak{b}_{2}}{\mathfrak{b}_{1}}\frac{(u-1)(u-2s-2)}{(u-S_{3}-s-1)(u-S_{3}-s-2)},\\
\mathbf{F}^{(2)}(u)= & \left[\mathbf{G}^{(2)}(u)\right]^{-1}\mathbf{G}^{(3)}(u)=\frac{1}{\mathfrak{b}_{2}^{2}}\frac{(u+S_{3}+s+1)(u-S_{3}-s-1)}{u^{2}}.
\end{split}
\end{equation}
The spectrum of the operators $S_{3}$ can be labeled by $m=-s,-s+1,\dots,s-1,s$,
therefore
\begin{equation}
\begin{split}\beta_{m} & =\mathfrak{b}_{1}^{J}\left(\frac{s+m+1}{2}\right)^{J},\\
\tilde{\mathcal{F}}_{m}^{(1)}(u) & =\frac{\mathfrak{b}_{2}}{\mathfrak{b}_{1}}\frac{(u+i/2)(u+2is+3i/2)}{(u+im+is+i/2)(u+im+is+3i/2)},\\
\mathcal{\tilde{\mathcal{F}}}_{m}^{(2)}(u) & =\frac{1}{\mathfrak{b}_{2}^{2}}\frac{u^{2}+(m+s+1)^{2}}{u^{2}}\sqrt{\frac{u^{2}}{u^{2}+1/4}}.
\end{split}
\end{equation}
The normalized overlap can be written as
\begin{multline}
\frac{\langle\mathrm{MPS}_{g}|\bar{u}\rangle}{\sqrt{\langle\bar{u}|\bar{u}\rangle}}=\mathfrak{b}_{1}^{J-r_{1}}\mathfrak{b}_{2}^{r_{1}-r_{2}}\sqrt{\frac{\det G^{+}}{\det G^{-}}}\times\\
\left[\sum_{m=-s}^{s}\left(\frac{s+m+1}{2}\right)^{J}\prod_{j=1}^{r_{1}}\frac{(u_{j}^{1}+i/2)(u_{j}^{1}+2is+3i/2)}{(u_{j}^{1}+im+is+i/2)(u_{j}^{1}+im+is+3i/2)}\prod_{j=1}^{r_{2}/2}\frac{(u_{j}^{2})^{2}+(m+s+1)^{2}}{u_{j}^{2}\sqrt{(u_{j}^{2})^{2}+1/4}}\right].
\end{multline}
In the $\mathfrak{b}_{1},\mathfrak{b}_{2}\to0$ limit the overlap
is non-vanishing only when $J=r_{1}=r_{2}$.

\section{Overlaps with odd number of chiral Bethe roots}

So far, for simpler presentation, we have focused on Bethe vectors
where each chiral nodus has an even number of Bethe roots. Small changes
are sufficient for generalization. In the following, we show how the
formulas are modified in the presence of odd numbers of chiral Bethe
roots.

When a chiral node $s$ have odd number of Bethe roots we define the
quantum numbers $r_{s}^{+}=\frac{r_{s}-1}{2}$ and $r_{s}^{0}=1$.
We also introduce the notations $u_{k}^{+,s}\equiv u_{k}^{s}$, $u_{k}^{-,s}\equiv u_{r_{s}+1-k}^{s}$
for $k=1,\dots,r_{s}^{+}$ and $u^{0,s}\equiv u_{r_{s}^{+}+1}^{s}$.
In the pair structure limit $u_{k}^{-,s}=-u_{k}^{+,s}$ for $k=1,\dots,r_{s}^{+}$
and $u^{0,s}=0$. When a node $s$ is achiral or chiral with even
number of Bethe roots then $r_{s}^{0}=0$. We also define the following
sets $\bar{u}^{\pm,s}=\left\{ u_{k}^{\pm,s}\right\} _{k=1}^{r_{s}^{+}}$,
$\bar{u}^{0,s}=\left\{ u^{0,s}\right\} _{k=1}^{r_{s}^{0}}$ and $\bar{u}^{\pm}=\left\{ \bar{u}^{\pm,1},\bar{u}^{\pm,2},\dots,\bar{u}^{\pm,n_{+}}\right\} $,
$\bar{u}^{0}=\left\{ \bar{u}^{0,1},\bar{u}^{0,2},\dots,\bar{u}^{0,n_{+}}\right\} $. 

The expressions $\phi_{k}^{s}$ can be grouped according to the pair
structure in the same way as the Bethe roots. According to the decompositions
$\bar{u}=\bar{u}^{+}\cup\bar{u}^{0}\cup\bar{u}^{-}$ and $\bar{\phi}=\bar{\phi}^{+}\cup\bar{\phi}^{0}\cup\bar{\phi}^{-}$
the Gaudin matrix has the following block form
\begin{equation}
G=\left(\begin{array}{ccc}
A^{++} & A^{+0} & A^{+-}\\
A^{0+} & A^{00} & A^{0-}\\
A^{-+} & A^{-0} & A^{--}
\end{array}\right).
\end{equation}
The Gaudin determinants are
\begin{equation}
\det G^{+}=\left|\begin{array}{cc}
A^{++}+A^{+-} & A^{+0}\\
A^{0+} & \frac{1}{2}A^{00}
\end{array}\right|,\quad\det G^{-}=2\left|A^{++}-A^{+-}\right|.\label{eq:degGodd}
\end{equation}
The overlap function is only minimally modified:
\begin{equation}
\frac{\langle\mathrm{MPS}|\bar{u}\rangle}{\sqrt{\langle\bar{u}|\bar{u}\rangle}}=\left[\sum_{k=1}^{d_{B}}\beta_{k}\prod_{j=1}^{n_{+}}\left(\gamma_{k}^{(j)}\right)^{r_{j}^{0}}\tilde{\mathcal{F}}_{k}^{(j)}(\bar{u}^{+,j})\right]\sqrt{\frac{\det G^{+}}{\det G^{-}}}.\label{eq:ov-1}
\end{equation}
We introduced a set of new parameters $\gamma_{k}^{(j)}\in\mathbb{C}$
for $k=1.\dots,d_{B}$ and chiral nodes $j$. These parameters can
be easily determined from numerical calculations.